\documentclass[jgrga]{agu2001}
\usepackage{graphicx}   

\authorrunninghead{PASQUALINI ET AL.}

\titlerunninghead{Nonequilibrium and Nonlinear Dynamics in 
Geomaterials}
\authoraddr{D. Pasqualini,
EES-9, University of California, Los Alamos 
National Laboratory, Los Alamos, New Mexico 87545, USA.
(dondy@lanl.gov)}

\begin{document} 

 \vbox to 0pt{\vss 
                    \hbox to 0pt{\hskip0pt\rm LA-UR-05-3335\hss} 
                   \vskip 25pt} 

\title{Nonequilibrium and Nonlinear Dynamics in Geomaterials I : The 
Low Strain Regime}

\author{Donatella Pasqualini, \altaffilmark{1}
Katrin Heitmann, \altaffilmark{2}
James A. TenCate, \altaffilmark{3}
Salman Habib, \altaffilmark{4}
David Higdon, \altaffilmark{5}
and Paul A. Johnson \altaffilmark{3}
} 
\altaffiltext{1}
{EES-9, University of California, Los Alamos 
National Laboratory, Los Alamos, New Mexico 87545} 
\altaffiltext{2}
{ISR-1, University of California, Los Alamos 
 National Laboratory, Los Alamos, New Mexico 87545} 
\altaffiltext{3}
{EES-11, University of California, Los Alamos 
National Laboratory, Los Alamos, New Mexico 87545} 
 \altaffiltext{4}
{T-8, University of California, Los Alamos 
 National Laboratory, Los Alamos, New Mexico 87545} 
\altaffiltext{5}
{D-1, University of California, Los Alamos 
 National Laboratory, Los Alamos, New Mexico 87545} 
  
\begin{abstract}
  Members of a wide class of geomaterials are known to display complex
  and fascinating nonlinear and nonequilibrium dynamical behaviors
  over a wide range of bulk strains, down to surprisingly low values,
  e.g., $10^{-7}$. In this paper we investigate two sandstones, Berea
  and Fontainebleau, and characterize their behavior under the
  influence of very small external forces via carefully controlled
  resonant bar experiments. By reducing environmental effects due to
  temperature and humidity variations, we are able to systematically
  and reproducibly study dynamical behavior at strains as low as
  $10^{-9}$. Our study establishes the existence of two strain
  thresholds, the first, $\epsilon_L$, below which the material is
  essentially linear, and the second, $\epsilon_M$, below which the
  material is nonlinear but where quasiequilibrium thermodynamics
  still applies as evidenced by the success of Landau theory and a
  simple macroscopic description based on the Duffing oscillator. At
  strains above $\epsilon_M$ the behavior becomes truly nonequilibrium
  -- as demonstrated by the existence of material conditioning -- and
  Landau theory no longer applies. The main focus of this paper is
  the study of the region below the second threshold, but we also
  comment on how our work clarifies and resolves previous experimental
  conflicts, as well as suggest new directions of research.
\end{abstract} 

\begin{article}

\section{Introduction}

\begin{figure}    
\begin{center}
\leavevmode\includegraphics[width=7cm,height=5cm]
{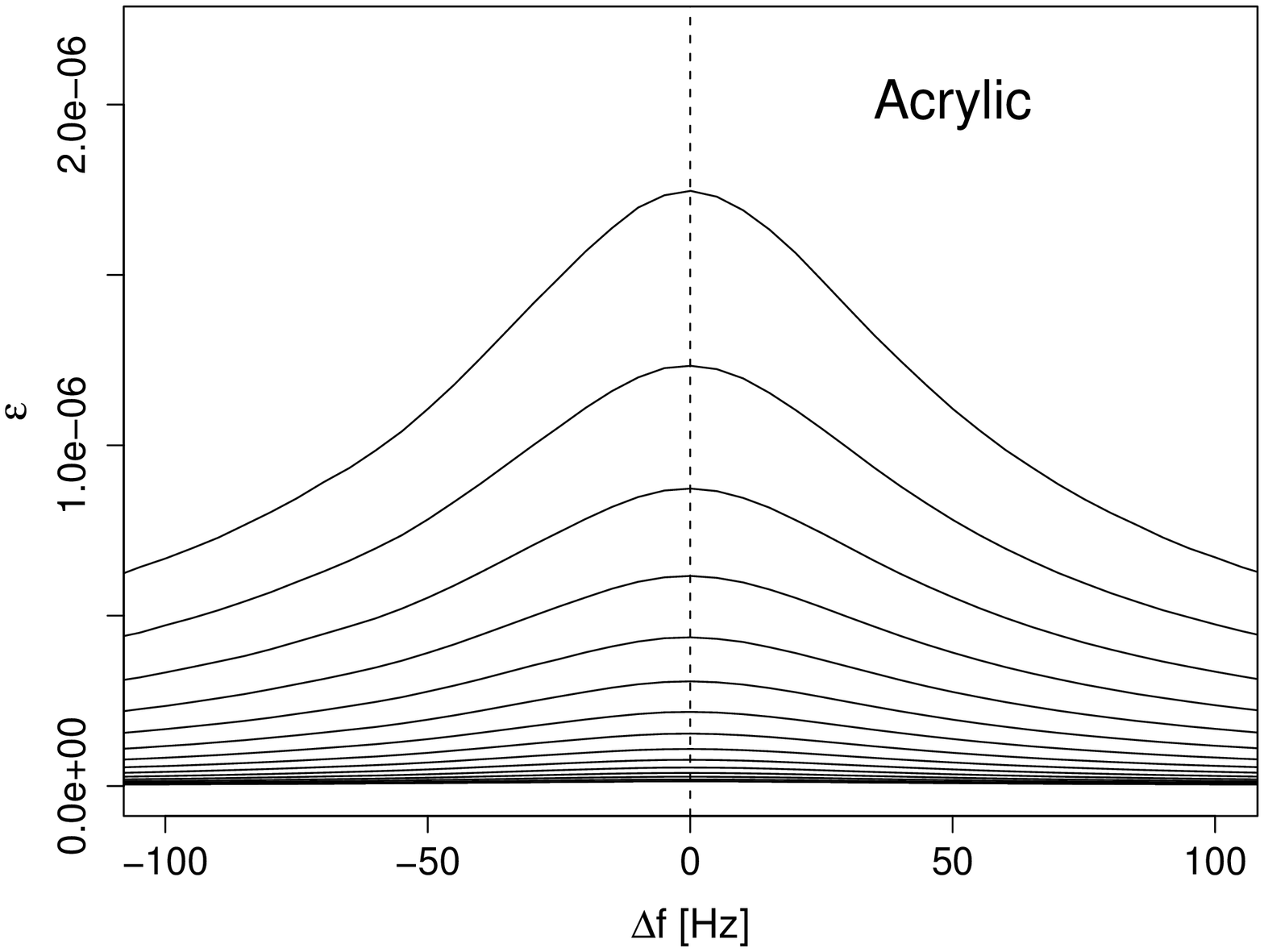}  

\vspace{0.5cm}

\leavevmode\includegraphics[width=7cm,height=5cm]
{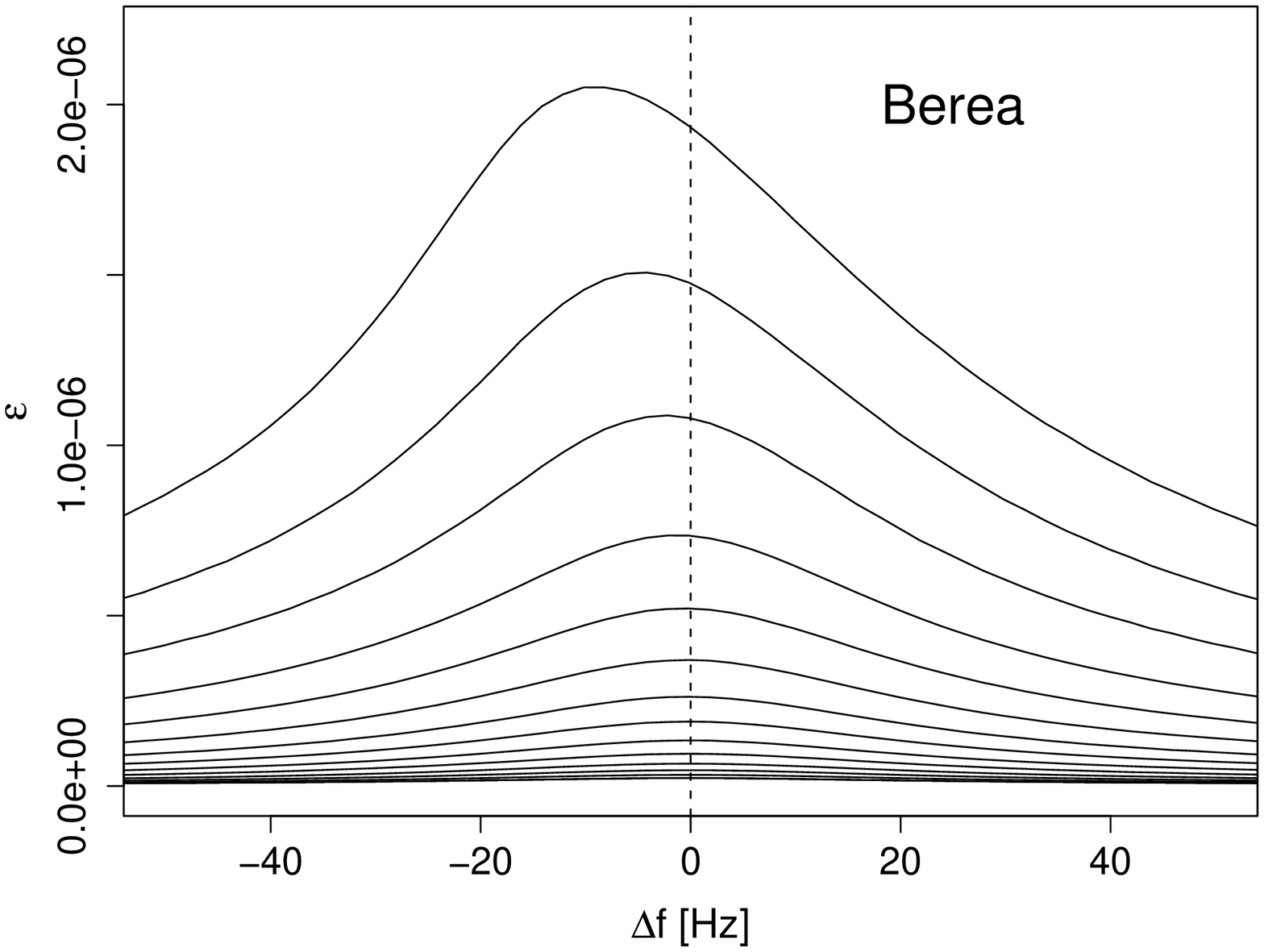} 

\vspace{0.5cm}

\leavevmode\includegraphics[width=7cm,height=5cm]
{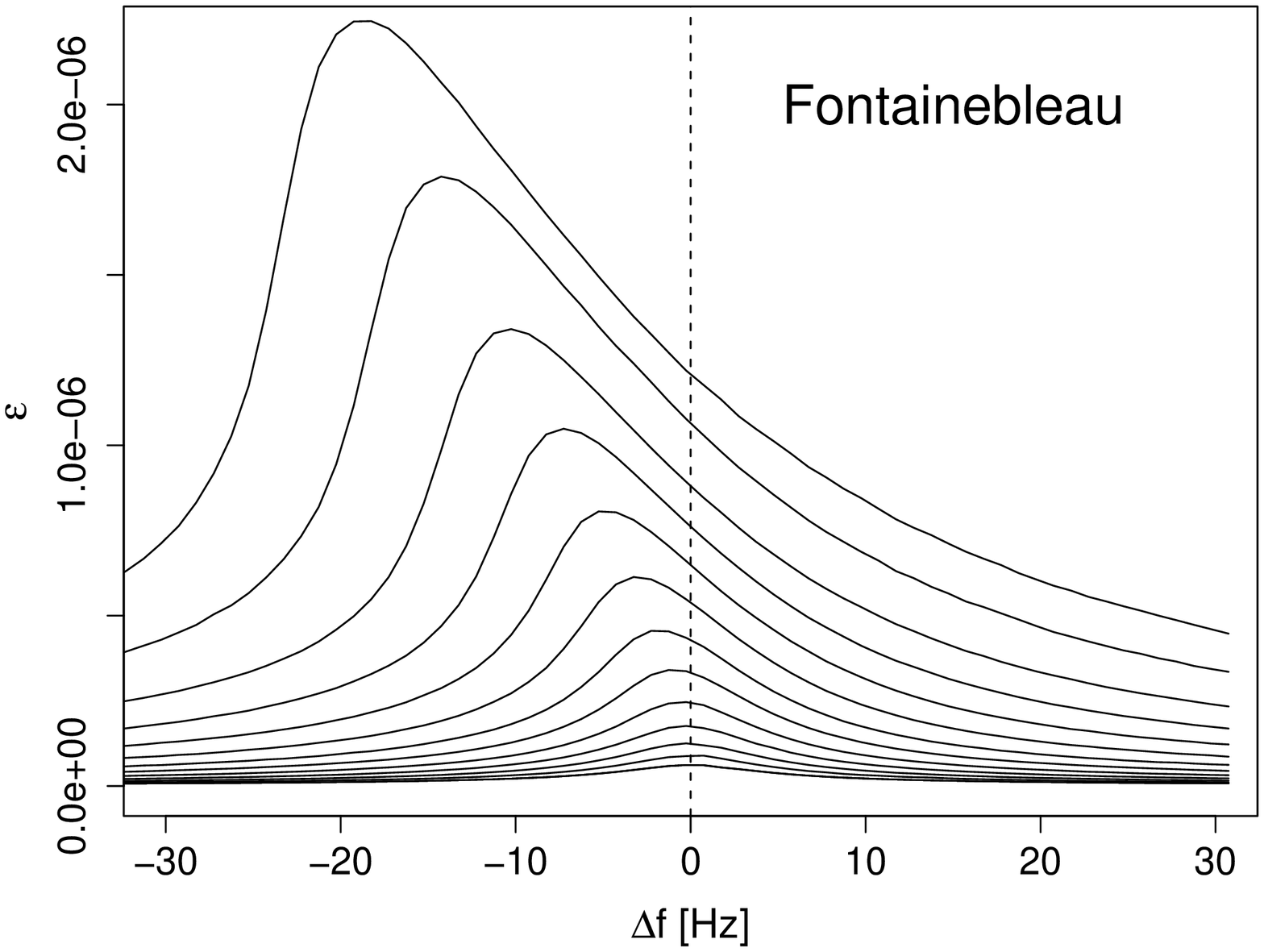}  
\end{center}

\caption{Resonance curves for Acrylic, Berea, and Fontainebleau at 
different drives. Acrylic is a linear material used as a control in 
the experiments. Nonlinearity is evidenced in Berea and Fontainebleau
samples by the shift in the peak of the resonance curves.  } 
\label{fig:RC} 
\end{figure}

Geomaterials display very interesting nonlinear features, diverse
aspects of which have been investigated over a long period of time
for a recent overview see, e.g., {\em Ostrovksy and Johnson},~2001
and references therein].  A standard technique used to study these
nonlinear features is the resonant bar experiment [{\em Clark}, 1966;
{\em Jaeger and Cook}, 1979; {\em Carmichael}, 1984; {\em Bourbie et
al.}, 1987].  In these experiments a long rod of the material under
test is driven longitudinally and its amplitude and frequency response
monitored.  For a {\em linear} material the resonance frequency of the
rod is invariant over a very wide range of dynamical strain. An
example of this behavior is shown in the results from one of our
experiments on Acrylic in the top panel of Figure~\ref{fig:RC}:
increasing the strain up to $2\cdot 10^{-6}$ leaves the resonance
frequency unchanged (note that the $x$-axis shows the change in the
resonance frequency, $\Delta f$, and not the resonance frequency
itself.)  The resonance frequency of a rod made from a {\em nonlinear}
material such as Berea sandstone behaves quite differently: When a
driving force is applied to the rod, the frequency either increases or
decreases (the modulus either hardens or softens) depending on the
precise properties of the material. This phenomenon is well-known and
a theoretical description based on quasiequilibrium thermodynamics and
nonlinear elasticity has existed for a long time [see e.g., {\em
Landau and Lifshitz}, 1998]; we will refer to this as the classical
theory of nonlinear elasticity or simply as Landau theory.

Many geomaterials, such as sandstones, belong to the general class of
nonlinear materials. The second and third panel in Figure~\ref{fig:RC}
display resonant bar results for two representative samples, Berea and
Fontainebleau.  In both cases the shift in the resonance frequency is
very large and the resonance frequency decreases with drive
amplitude. The strength of the nonlinear response in these materials
is very large, orders of magnitude more than for metals. Consequently,
it is important to check whether Landau theory still applies to these
materials, and, if so, over what range of strains.

It is widely believed that geomaterials behave differently than weakly
nonlinear materials because of their complex internal structure.  They
are formed by an assembly of more or less rigid ``grains'' connected
via a much softer ``bond'' network of varying porosity.  The grains
make up a large fraction of the volume, between 80 and 99\%.
Individual grains can be very pure (as in the case of Fontainebleau,
$\sim 99+\%$ quartz) or made up from several different components (as
in the case of Berea: 85\% quartz, 8\% feldspar, plus small quantities
of other minerals).  Most of these materials are quite porous and
their behavior changes dramatically under the influence of
environmental effects, such as temperature [see e.g. {\em Sheriff},
1978] or humidity [see e.g., {\em Gordon and Davis}, 1968; {\em
O'Hara}, 1985; {\em Zinszner et al.}, 1997; {\em Van den Abeele et
al.}, 2002].  This sensitivity to the environment makes controlled
studies difficult, as the experiments must be carried out in such a
way that these effects are demonstrably under control.

Another difficulty in measuring the frequency response of sandstones
arises from the brittleness of rocks. If the samples are driven too
hard, microcracks can be induced and the resulting behavior of the
material can change dramatically. In addition, driving can also induce
long-lived nonequilibrium macrostates that relax back over a long
period of time ($\sim$ hours). Thus, it is important to ensure -- by
repeating a given drive protocol on the same sample and verifying that
the material response does not change from one experiment to the next
-- that the samples have not been altered from their original
condition and the environment is unchanged over the set of
observations. The experiments described in this paper were carried out
in this way. Furthermore, the very low strain values ensured that
sample damage rarely occurred.

One goal of this work is to clarify, using new and existing data,
conflicting observations in the literature, and to present a
description of the ``state of the art'' at low strain amplitudes. Here
we restrict ourselves mainly to the question of dynamic nonlinearity
and do not take up the equally important question of the nature of
loss mechanisms and their connection and interaction with the
nonlinear (compliant) behavior underlying the frequency shift.

In the past, several different groups have carried out resonant bar
experiments.  {\em Gordon and Davis}~[1968] investigated a large suite
of crystalline rocks, including Quartzite, Granite, and Olivine
basalt, at strains between $10^{-9} < \epsilon < 10^{-3}$.  Their main
objective was to measure the loss factor $Q^{-1}$ (or the internal
friction $\phi$ in their terminology) as a function of strain and the
ratio of stress and strain.  In order to cover the large strain range
they divided their experiments in two components: for $10^{-9} <
\epsilon < 10^{-5}$ they used the driven frequency method, driving the
rocks at very high frequencies, and for $10^{-5} < \epsilon < 10^{-3}$
they made direct measurements of the stress-strain curve.  Their main
findings are the following.  (i) The loss factor is quite insensitive
to the strain amplitude, diverging from a constant value only at high
strains.  At these high strains they conclude that this increase in
$Q^{-1}$ is the result of internal damage.  (ii) $Q^{-1}$ is highly
structure sensitive, i.e., it is sensitive to the details of the
microstructure of the rock.  (iii) $Q^{-1}$ increases as the
temperature increases.  They conclude that this increase is due to
grain-interface displacement, and therefore alteration of the internal
structure of the rock.  (iv) At large strains they find static
hysteresis with end-point memory.

Following up on {\em Gordon and Davis} [1968], {\em McKavanagh and
Stacey} [1974] and {\em Brennan and Stacey} [1977] performed another
set of stress-strain loop measurements on granite, basalt, sandstone,
and concrete.  Their main objective was the measurement of
stress-strain loops below strain amplitudes of $\epsilon = 10^{-5}$,
since {\em Gordon and Davis} [1968] had reported that $Q^{-1}$ above
this limit was no longer a linear function of the applied strain.
{\em McKavanagh and Stacey} [1974] were able to go down to strains of
$10^{-6}$.  (Note that this level is still {\em above} the strain at
which we found nonequilibrium effects to be important, {\em TenCate et
al.}  [2004].)  At these strains they found that the hysteresis loops
for sandstone were always cusped at the ends.  Another interesting
result was that below a certain strain amplitude the shape of the loop
became independent of the applied strain amplitude. From this they
concluded that even at the very smallest strain amplitudes, cusps
should continue to be present in stress-strain loops. (However, {\em
Brennan and Stacey} [1977] noted that for granite and basalt, the
stress-strain loops do become elliptical for strains lower than
$10^{-6}$.) In view of our recent results [{\em TenCate et al.}, 2004]
this conclusion might have been drawn without having enough evidence
at low enough strain amplitudes.  We return to this point later in
Section~\ref{stressstrainloop}.

{\em Winkler et al.}~[1979] conducted experiments with Massilon and
Berea sandstone at strain amplitudes between 10$^{-8}$ and 10$^{-6}$.
The main goal was to determine the strains at which seismic energy
losses caused by grain boundary friction become important but
softening of the resonance frequency with strain amplitude was also
investigated.  They concluded that the losses are only important at
strains larger than were investigated.  Additionally, they found that
the two sandstones investigated displayed nonlinear features dependent
on several external parameters, such as water content or confining
pressure.  They find that the loss factor is independent of strain
below strains of $5\cdot 10^{-7}$ while at relative large strain ( $>
10^{-6}$) there is a clear increase, in agreement with {\em Gordon and
Davis} [1968].  The main drawback of the experiments by {\em Winkler
et al.}  [1979] is the relative lack of data points, especially in the
very low strain regime; the quality of the repeatability of their
measurements on the same sample is also not shown.  In this respect,
our work significantly improves on previous results; we increase the
number of measurement points in the low strain regime by a factor of
five in comparison to {\em Winkler et al.} [1979], allowing a more
robust analysis of the data.

More recently, {\it Guyer et al.}~[1999] and {\it Smith and
TenCate}~[2000] analyzed a set of resonant bar experiments with Berea
sandstone samples also at low strains.  The conclusions they reached,
however, were in strong disagreement with the older results of, e.g.,
{\it Winkler et al.}~[1979].  Instead of the expected quadratic
behavior of the frequency shift with drive at very low strains -- an
essential prediction of Landau theory -- they reported an ostensibly
{\it linear} dependence, claimed to hold down to the smallest strains.
We note that such a linear softening in several material samples was
also reported in {\em Johnson and Rasolofosaon} [1996] (see also
references therein), albeit at significantly higher strains.

This surprising behavior was claimed to be consistent with predictions
of a phenomenological model originally developed to explain (static)
hysteretic behavior in geomaterials at very high strains [the
Preisach-Mayergoyz space (PM space) model].  In this model a rock
sample is described in terms of an ensemble of mesoscale hysteretic
units [{\em McCall and Guyer}, 1994; {\em Guyer et al.}, 1997].  By
applying the PM space model to low-strain regimes, a linear dependence
of the frequency shift with drive can be obtained.  By its very
nature, the model also predicts the existence of cusps in
low-amplitude stress-strain loops.  As discussed in
Section~\ref{stressstrainloop}, however, we do not detect cusps in
stress-strain loops at low strains.

Motivated partly by these very different findings on similar
sandstones and with similar experimental set-ups, we embarked on a set
of well-characterized resonant-bar experiments using Fontainebleau and
Berea sandstone samples {\it TenCate et al.}~[2004]. Broadly speaking,
our findings for the resonance frequency shift confirm the original
results of {\it Winkler et al.}~[1979]; below a certain strain
threshold $\epsilon_M$ both sandstones displayed the expected
quadratic behavior. In addition, we were able to show that previous
claims of a linear shift at high strains are actually an artifact due
to the material conditioning mentioned above at strains higher than
$\epsilon_M$, and that a simple macroscopic Duffing model provides an
excellent mathematical description of the experimental data without
going beyond Landau theory (as PM-space models explicitly do, by
adding nonanalytic terms to the internal energy expansion). Thus, we
established that, to the extent macro-reversibility holds, the
predictions of classical theory are in fact correct.

In this paper we extend our previous analysis by adding an
investigation of energy loss (via the resonator quality factor $Q$),
dynamical stress-strain loops, and harmonic generation. We carry out
the same experiment several times with the same sample to demonstrate
environmental control and repeatability.  The data analysis is based
on a Gaussian process model to avoid biasing from nonoptimal fitting
procedures applied to experimental data. The Duffing model introduced
in our previous work is shown to be nicely consistent with the newer
results. The predictions of this model for the quality factor, the
frequency shift, and hysteresis cusps (null prediction) all hold
within experimental error at strains below $\epsilon_M$. At higher
strains, this simple model breaks down -- as it must -- due to the
(deliberate) exclusion of nonequilibrium effects.  Finally, we have
reanalyzed a subset of the data which were taken in 1999 [{\it Smith
  and TenCate},~2000] and had led to very different conclusions for
Berea samples. We show that the interpretation of the data in the
earlier papers was incorrect and demonstrate that the experimental
data are actually in good agreement with our present findings [this
paper and {\em TenCate et al.}, 2004].

The paper is organized as follows.  First, in
Section~\ref{section:experiments} we describe the experimental set-up
in some detail. Next, in Section~\ref{section3} we explain how we
analyze the data, especially how we determine the peaks of the
resonance curves and how our procedure allows us to determine
realistic error bars. In Section~\ref{expres} we discuss the results
from the experiments. A simple theoretical model that describes the
experimental results is presented in Section~\ref{section:model}.  We
confront previous findings in very similar experiments with our new
results in Section~\ref{section:old} and conclude in
Section~\ref{conclusions}.

\section{Experiments}\label{section:experiments}

%Samples
The samples used in the experiments are thin cores of Fontainebleau
and Berea sandstone\footnote{Sources: Fontainebleau: IFP, Berea:
  Cleveland Quarz Ohio}, 2.5~cm in diameter and 35~cm long.  As
established by X-ray diffraction measurements, the Fontainebleau
sandstone is almost pure quartz ($>$99\% with trace amounts of other
materials); Berea sandstone is less pure having only 85$\pm$8\% quartz
with 8$\pm$1\% feldspar and 5$\pm$1\% kaolinite and approximately 2\%
other constituents.  Fontainebleau sandstone has grain sizes of around
150~$\mu$ and a porosity of $\approx$~24\%.  Berea sandstone samples
have grain sizes which are somewhat smaller, $\approx$~100~$\mu$, with
a porosity of about 20\%.

%Receiver/Sources
A small Bruel\&Kj{\ae}r~4374 accelerometer is carefully bonded to one
end of each core sample with a cyanoacrylate glue (SuperGlue gel,
Duro).  The accelerometers are an industry standard, and are well
characterized.  With perfect bonding between accelerometer and rock,
the accelerometer -- and the associated B\&K 2635 Charge Amp -- has a
flat frequency and phase response to 25~kHz.  With poor bonds, the
upper frequency limit of the flat response drops.  Thus, great care is
taken to establish a good bond between accelerometer and sample.  Each
accelerometer is first qualitatively tested (i.e., finger pressure) to
be sure of a strong bond. Furthermore, before the samples are placed
in the environmental isolation chamber (discussed below) for
measurements, a comparison of the accelerometer response with a laser
vibrometer (Polytec) is made and accelerometers are rebonded if the
frequency responses differed noticeably. In any case, it is important
to point out that for the samples used in this study, all of the
resonance frequencies are below 3~kHz, nearly an order of magnitude
below the upper frequency flat response limit for the
accelerometer/charge amp combination.

The source excitation is provided by a 0.75~cm thick piezoelectric
disk epoxied (Stycast~1266) to the other end of the sample core and
backed with an epoxied high impedance backload (brass) to ensure that
most of the acoustic energy couples into the rock sample instead of
the surrounding environment.  Resonances in the backload ($>~$50 kHz)
are much higher than the frequencies and resonances of the sample and
thus are not excited in our experiments.

\begin{figure}
\begin{center}
\leavevmode\includegraphics[width=7cm,height=5.5cm]
{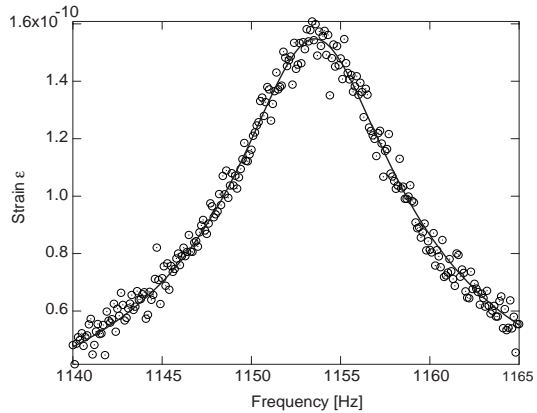}
\end{center}
\caption{Low-amplitude drive resonance curve for Fontainebleau
  sandstone. The solid curve is a Lorentzian fit to the data points.}  
\label{fig:exp}
\end{figure}

%Longitudinal mode resonance experimental comments
For all the experiments described here, the lowest order longitudinal
mode (the first Pochhammer mode) is excited.  (We note that the mass
of the brass backload lowers the center frequencies of the Pochhammer
mode resonances somewhat but does not affect the shape of a resonance
curve.)  Resonance curves are easy to measure and analyze and fairly
high strains can be attained without requiring a high-power amplifier
(with its frequently accompanying nonlinearities).  For the
Fontainebleau sandstone the lowest resonance frequency is around
1.1~kHz; for the Berea sandstone the lowest resonance frequency is
around 2.8~kHz.  Measured values for the quality factor $Q$ of these
resonances are about 130 for the Fontainebleau sandstone sample and
about 65 for the Berea sandstone sample.  The lowest order Pochhammer
mode has both compressional and shear components but the motion is
nevertheless quasi-one-dimensional and the bulk of the sample
participates in the wave motion associated with the resonance.  As
higher-order Pochhammer modes begin to resemble surface waves, only
the very lowest frequency modes are examined here.

The samples are suspended at two points with loops of synthetic fiber
(dental floss) or thin O-rings.  Different suspension points slightly
alter the lowest Pochhammer mode resonance frequencies but these
differences are much smaller than differences caused by even slight
changes of temperature; moreover, and perhaps more importantly, once
the bar is mounted, the resonance frequencies do {\it not} change with
increasing drive levels when tested with a standard (an acrylic bar).
Suspended in this way (stress-free ends) the sample's lowest
Pochhammer resonance frequency corresponds to roughly a
half-wavelength in the sample.

%Environmental chamber description.
Since most rocks are extremely sensitive to temperature and
temperature {\it changes} [{\it Ide}, 1937] -- with relaxation times
of several hours -- we have built a sample chamber for effective
environmental isolation.  An inner 3/4-inch-wall plexiglass box with
caulked seams holds both the samples which are suspended from the top
of the box.  Air-tight electrical feedthroughs are available for
driver and accelerometer connections.  The entire chamber is flushed
with N$_2$ gas and then placed inside another (larger) plexiglass box
and surrounded with fiberglass insulation and sealed.  The inner
sample chamber also sits on top of gel pads for vibration
isolation. The complete isolation chamber is placed in a room whose
temperature is controlled with a thermostat and typically varies by no
more than 3 degrees C. Measured resonance frequencies of samples in
this box have been stable to within 0.1~Hz.

%Measurement electronics
To get the most precise measurements possible, we use an HP~3325B
Frequency synthesizer with a crystal oven for frequency stability as
the signal source.  The signal from the HP~3325B is fed into the
reference input of an EG\&G 5301A Lock-In amplifier which compares
that reference signal with the measured signal from the accelerometer
via a B\&K~2635 charge amplifier.  The whole experiment, including
data acquisition, is computer controlled via LabVIEW and a GPIB bus.
To drive the source, the signal from the HP frequency synthesizer is
fed into a Crown Studio Reference I amplifier and matched to the
(purely capacitive) piezoelectric transducer via a carefully
constructed and tested linear matching transformer.

%Testing of whole measurement system (standards and stability)
To test all the electronics for linearity, we have constructed several
known linear sample standards of nearly identical geometry to the rock
samples.  The density, sound speed, and $Q$'s of the samples are
chosen such that the mechanical impedances $\rho\cdot c$ are similar
to those of the rock samples.  These ``standard'' samples are driven
with identical source/backloads and at levels similar to those
experienced by the rock samples.  No nonlinearities have been seen;
results for an acrylic rod are shown in Figure~1.

%Testing on a real sample, how much better can we do than previously?
With the present isolation system, we have verified long-term
frequency stability of the samples to $\pm$ 0.1 Hz (corresponding to a
long-term thermal stability inside the chamber of 10~mK), which is
close to how well the peak of the frequency response curve can be
determined at the lowest levels of strain shown in this paper.  To
test the sensitivity of the Lock-In amplifier and assembled apparatus,
we have measured a resonance curve on the Fontainebleau sample at an
extremely low drive level.  The result is shown in Figure~\ref{fig:exp}.
The acceleration measured by the accelerometer has been converted to
strain (the open circles) using the driving frequency $f$ via
$\epsilon=\ddot{u}/(4\pi L f^2)$ following the convention in {\it
  TenCate et al.}~[2004].  Even though the peak strain near the
resonance frequency is only about 1.6 $\cdot 10^{-10}$, the shape of
the resonance curve is clear with only minimal noise obscuration: a
Lorentzian curve is an extremely good fit to the data as shown by the
solid line.  (Error bars are not shown for clarity.)  With computer
control and long-term temperature stability due to the isolation
chamber, this experimental setup permits long enough times to take
data over a large -- and an order of magnitude lower -- range
of strains not studied previously.

\section{Data Analysis}\label{section3}

\begin{figure}
\begin{center}  
\leavevmode\includegraphics[width=7cm,height=5.5cm]
{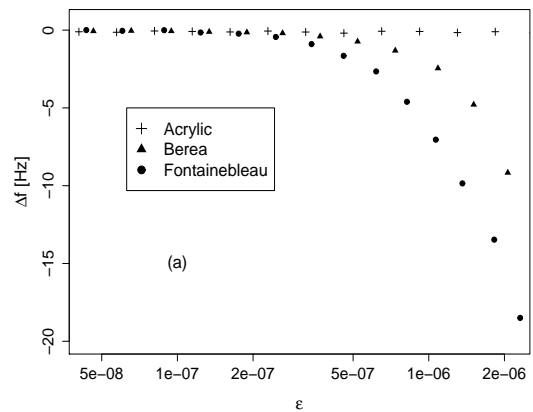} 
\end{center}
\caption{Resonance frequency shift $\Delta f$ as a function of the 
effective strain $\epsilon$ for the three samples shown in 
Figure~\ref{fig:RC}. } 
\label{fig:allrange_shift} 
\end{figure}

The basic quantities measured in a resonance experiment are the
frequency $f$ and the accelerometer voltage $V$, which is
automatically converted into acceleration $\ddot{u}$.  It is
convenient to translate the acceleration to a strain variable in order
to make the comparison of different samples with different lengths
easier.  As stated earlier, we employ the convention $\epsilon =
\ddot{u}/(4\pi Lf^2 )$, where $L$ is the length of the bar.  These
measurements lead to resonance curves as shown e.g., in
Figure~\ref{fig:RC}.  The task now is to determine the peaks of the
resonance curves, tracking the shift of the resonance frequency as a
function of the strain as displayed in
Figure~\ref{fig:allrange_shift}.

%Other people's methods, e.g., Winkler?
In the past, different methods have been suggested to analyze data
from low-strain resonant bar experiments [earlier attempts include
{\it Guyer et al.},~1999; {\it Smith and TenCate},~2000]. In this
paper we use a statistical analysis based on a nonparametric Gaussian
process to model the strain $\epsilon$ as a function of the driving
frequency $f$.  The flexibility of the Gaussian process model for
strain allows for estimation of the resonance frequency and resulting
strain $(f^*,\epsilon^*)$ without assuming a parametric form for the
dependence of strain on driving frequency.  Drawbacks of using a
parametric model can include understated uncertainties regarding
resonance quantities $(f^*,\epsilon^*)$ and excessive sensitivity to
measurements far away from the actual resonance frequency.  The
nonparametric modeling approach avoids both of these possible
pitfalls.

For a given experiment, observations $(f_i,\epsilon_i),\;i=1,\ldots,n$
are taken.  The observed strain is modeled as a smooth function of
frequency plus white noise $\delta$:
\begin{equation}
\epsilon_i = z(f_i) +
\delta_i,\,i=1,\ldots,n, 
\end{equation}
where the smooth function $z(f)$ is modeled as a Gaussian process and
each $\delta_i$ is modeled as an independent $N(0,\sigma^2)$ deviate.
The Gaussian process model for $z(f)$ is assumed to have an unknown
constant mean $\mu$ and a covariance function of the form
\begin{equation}
C[z(f_i),z(f_j)] = \sigma^2_z
\rho^{-|f_i - f_j|^2}.  
\end{equation}
The model specification is completed by specifying prior distributions
for the unknown parameters $\sigma^2$, $\mu$, $\sigma^2_z$, and $\rho$.
After shifting and scaling the data so that the $f_i$'s are between 0
and 1, and the $\epsilon_i$'s have mean 0 and variance 1, we fix $\mu$
to be 0 and assign uniform priors over the positive real line to
$\sigma^{-2}$ and $\sigma^{-2}_z$, and a uniform prior over [0,1] to
$\rho$.

\begin{figure}[t]
\begin{center}  
\leavevmode\includegraphics[width=7cm,height=5.5cm]
{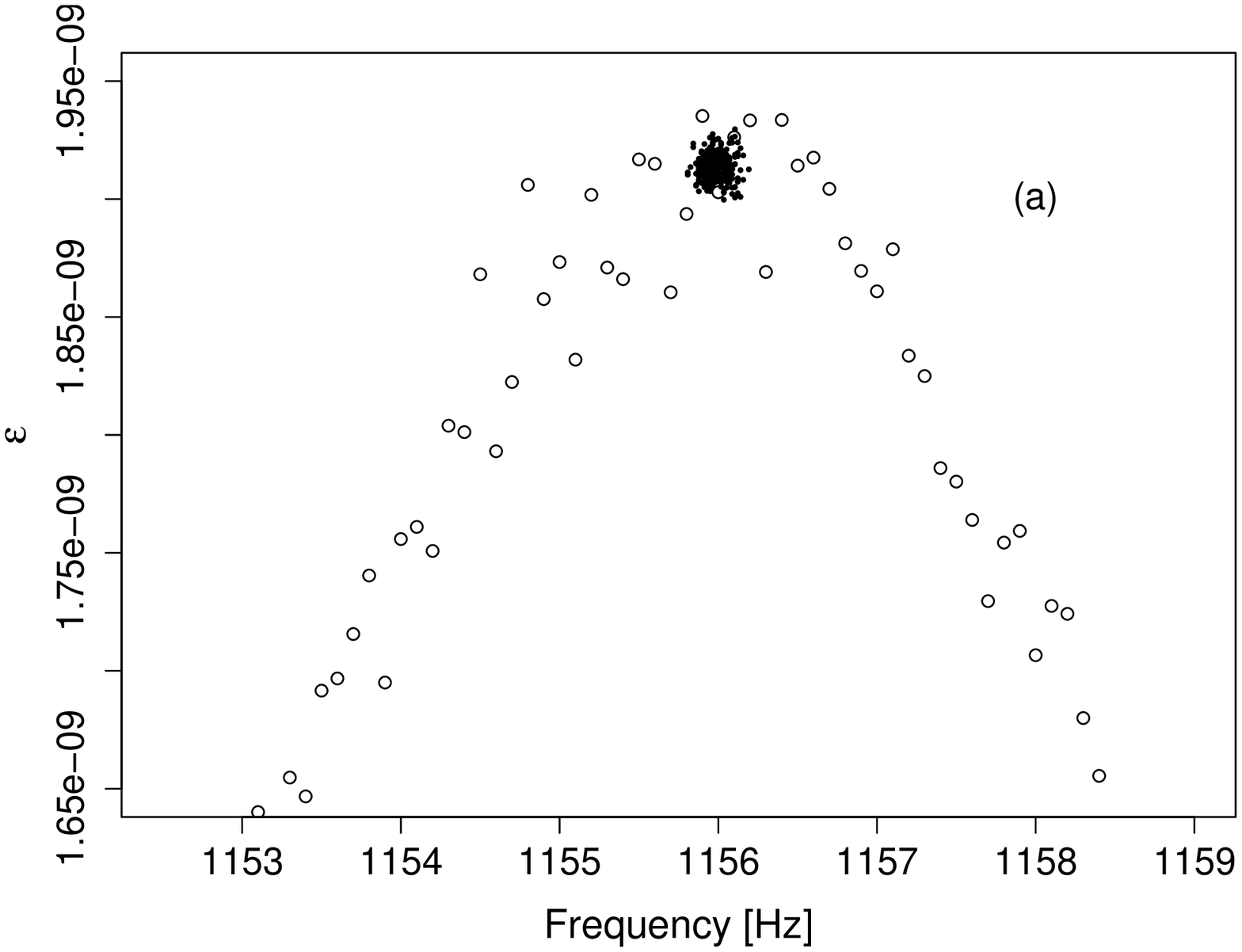} 

\vspace{0.5cm}

\leavevmode\includegraphics[width=7cm,height=5.5cm]
{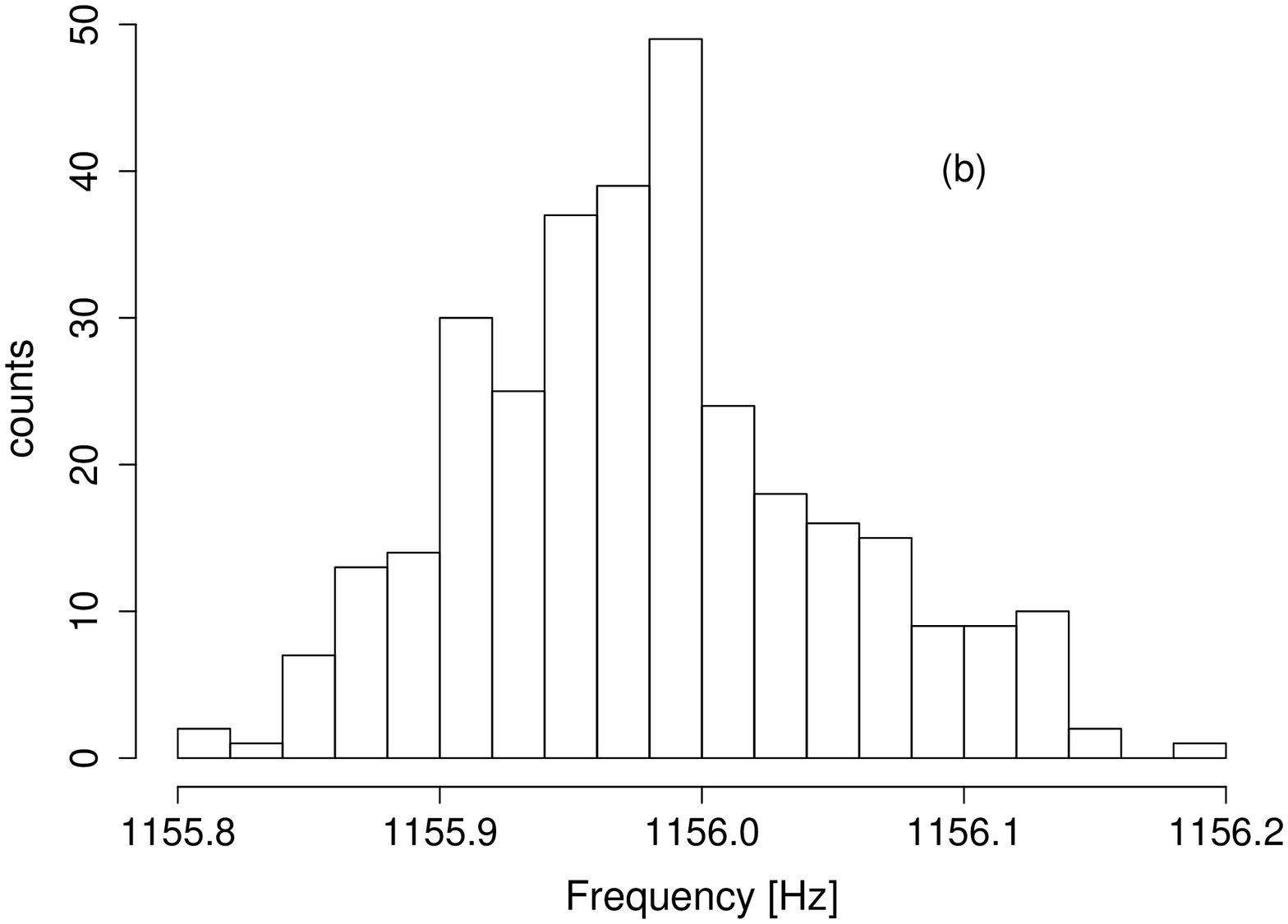}
  
\vspace{0.5cm}

\leavevmode\includegraphics[width=7cm,height=5.5cm]
{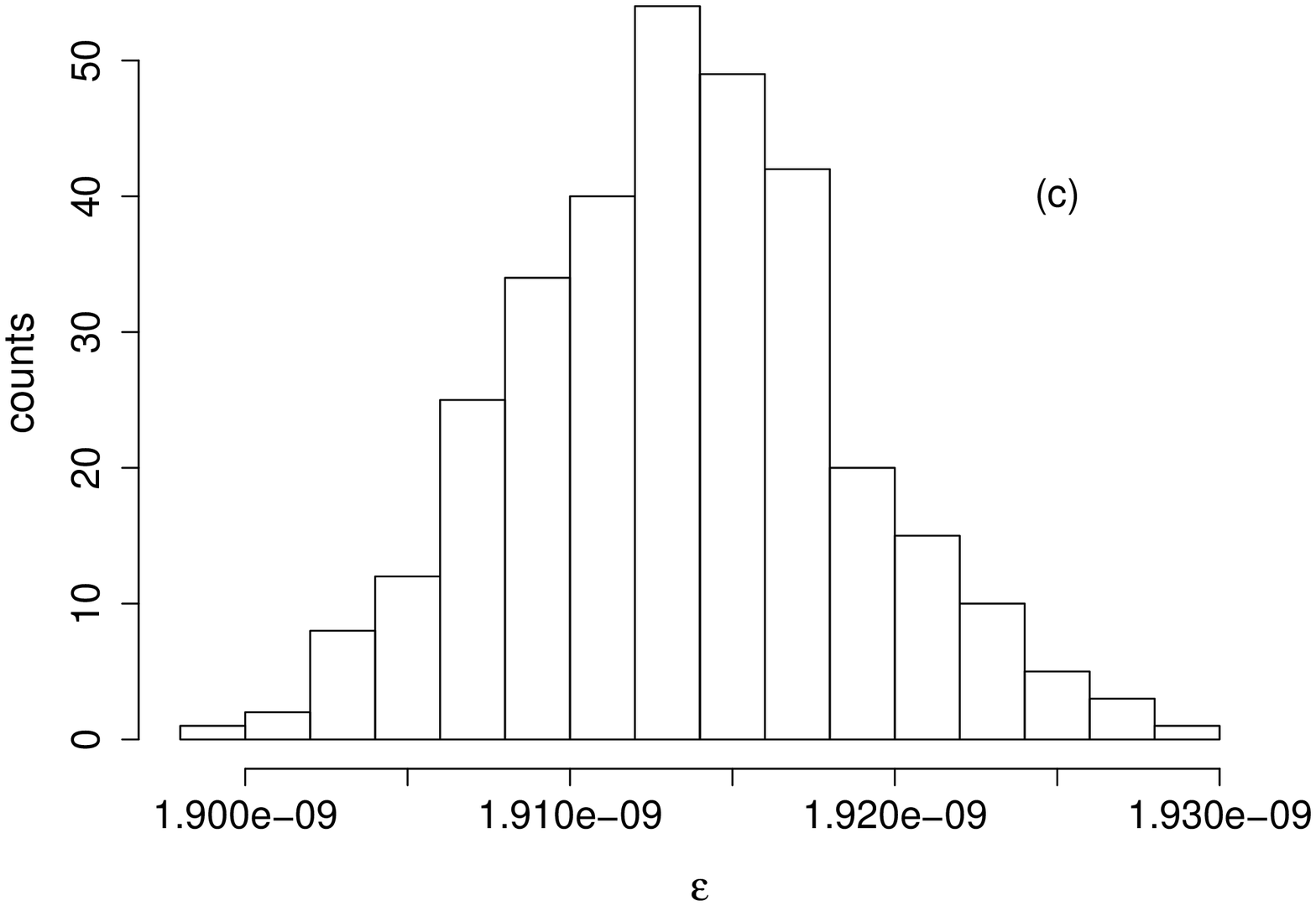}

\end{center}
\caption{(a) Resonance curve for Fontainebleau at a strain
  $\sim$2$\cdot 10^{-9}$. The central cluster of dots is the MCMC 
  posterior sample of pairs (f$^*$, $\epsilon^*$) that define the
  resonance peak. Frequency peak distribution (b) and frequency peak
  strain distribution (c) from the MCMC analysis for the same
  resonance curve shown in (a).}
\label{fig:mcmc}
\end{figure}

The resulting analysis gives a posterior distribution for the unknown
function $z(f)$ which we take to be the resonance curve.  This
posterior distribution quantifies the updated uncertainty about $z(f)$
given the experimental observations.  We use a Markov chain Monte
Carlo (MCMC) approach to sample realizations from the posterior
distribution of $z(f)$ over a dense grid of points in the neighborhood
of the resonance frequency $f^*$ [{\it Banerjee et al.},~2004].  From
each of these MCMC realizations of $z(f)$ the resonance frequency
$f^*$ and the corresponding maximum strain $\epsilon^* = z(f^*)$ are
recorded.  This creates a posterior sample of pairs $(f^*,\epsilon^*)$
which are given by the dots in Figure~\ref{fig:mcmc}(a).
Figures~\ref{fig:mcmc}(b) and \ref{fig:mcmc}(c) show the posterior
uncertainty for $f^*$ and $\epsilon^*$ separately with histograms of
these posterior samples.  We use the posterior mean as point estimates
for $f^*$ and $\epsilon^*$.  Later in the paper we use error bars that
connect the 5th and the 95th percentiles of the posterior samples to
quantify the uncertainty in our estimates.

\section{Experimental Results}
\label{expres}

\subsection{Memory Effects and Conditioning}
We have recently established the existence of two strain regimes [{\it
TenCate et al.},~2004]. As mentioned earlier, in the first regime
(strains below $\epsilon_M$) the material displays a {\em reversible}
softening of the resonance frequency with strain, while in the second
regime, (nonequilibrium) memory and conditioning effects become
apparent.  The second regime is entered at the strain threshold
$\epsilon_M$ which depends on the material and the environment (e.g.,
temperature, saturation etc.).  To determine $\epsilon_M$ for these
samples, the following experiments are performed.

\begin{figure}[b]
\begin{center}  
\leavevmode\includegraphics[width=7cm,height=5.5cm]
{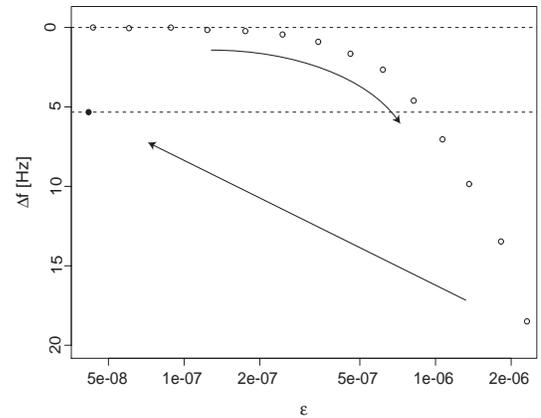} 
\end{center}
\caption{Example of resonance frequency shift showing the
  conditioning effect. The drive is increased up to a strain of
  $2\cdot$$10^{-6}$ and afterwards the rock is driven again at the
  lowest strain. The black dot shows the value of the resonance
  frequency peak after the last drive application. The difference
  between the two values for $\Delta f$ at the lowest strain
  demonstrates the effect of conditioning.} 
\label{fig:expShift} 
\end{figure}

A reference resonance curve is obtained at the lowest strain possible.
The resonance frequency is determined and used as a reference
frequency $f_0$ for the following procedure.  The source excitation
level is increased, a new resonance curve is obtained, and then
followed immediately by dropping the excitation level back in an
attempt to repeat the reference resonance curve.  If there are no
memory effects, the repeated curve's resonance frequency should match
the initial reference frequency.  If memory effects are at play, they
will persist and the repeated curve's peak resonance frequency will be
lower than the original.  An example of this is shown in
Figure~\ref{fig:expShift}.  This procedure is repeated for
incrementally increasing excitation levels until memory effects become
measurable.  The excitation level (and strain) where memory effects
first become noticeable defines $\epsilon_M$ for that sample.

The existence of the two regimes delineated by $\epsilon_M$ is crucial
to understanding and interpreting the dynamical behavior of
geomaterials.  Although it is possible to describe the nonlinearity of
the material at strains below $\epsilon_M$ with classical theory [{\it
Landau and Lifshitz},~1998], above $\epsilon_M$ the experimental
results are complicated by conditioning effects due to the
nonequilibrium dynamics of the rock.  Disentangling the intrinsic
nonlinearity of the material and these nonequilibrium effects is very
difficult and the frequency shifts in dynamical experiments at strains
above $\epsilon_M$ do not have a simple interpretation.  In
particular, classical elasticity theory assumes thermodynamic
reversibility and therefore cannot be applied in this essentially
nonequilibrium situation.  By the same token, {\em classical theory
cannot be tested by experiments carried out in this regime}.  As
discussed in the Introduction, previous experimental data were
interpretated without properly taking the existence of these different
regimes into account [e.g.,~{\it Guyer and Johnson},~1999].  This,
along with incorrect analysis of the experimental data (see the
discussion below), led to claiming evidence for nonclassical behavior
where in fact none existed.  Nevertheless, it is clear that a new
theoretical framework for the second regime, one that combines
nonlinearity with nonequilibrium dynamics, is definitely needed.

\begin{figure}  
\begin{center}
\leavevmode\includegraphics[width=7cm,height=5.5cm]
{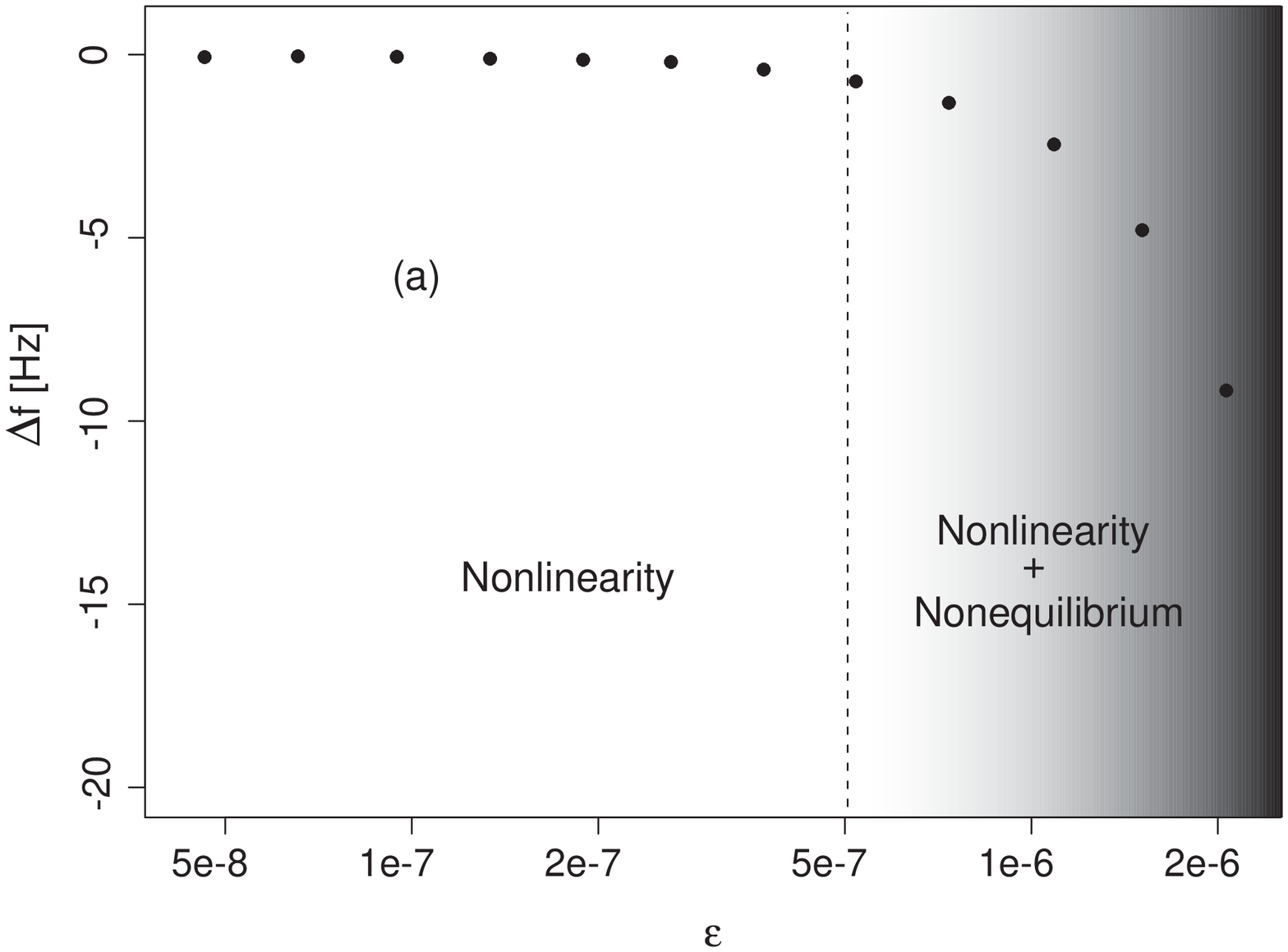} 

\vspace{0.5cm}

\leavevmode\includegraphics[width=7cm,height=5.5cm]
{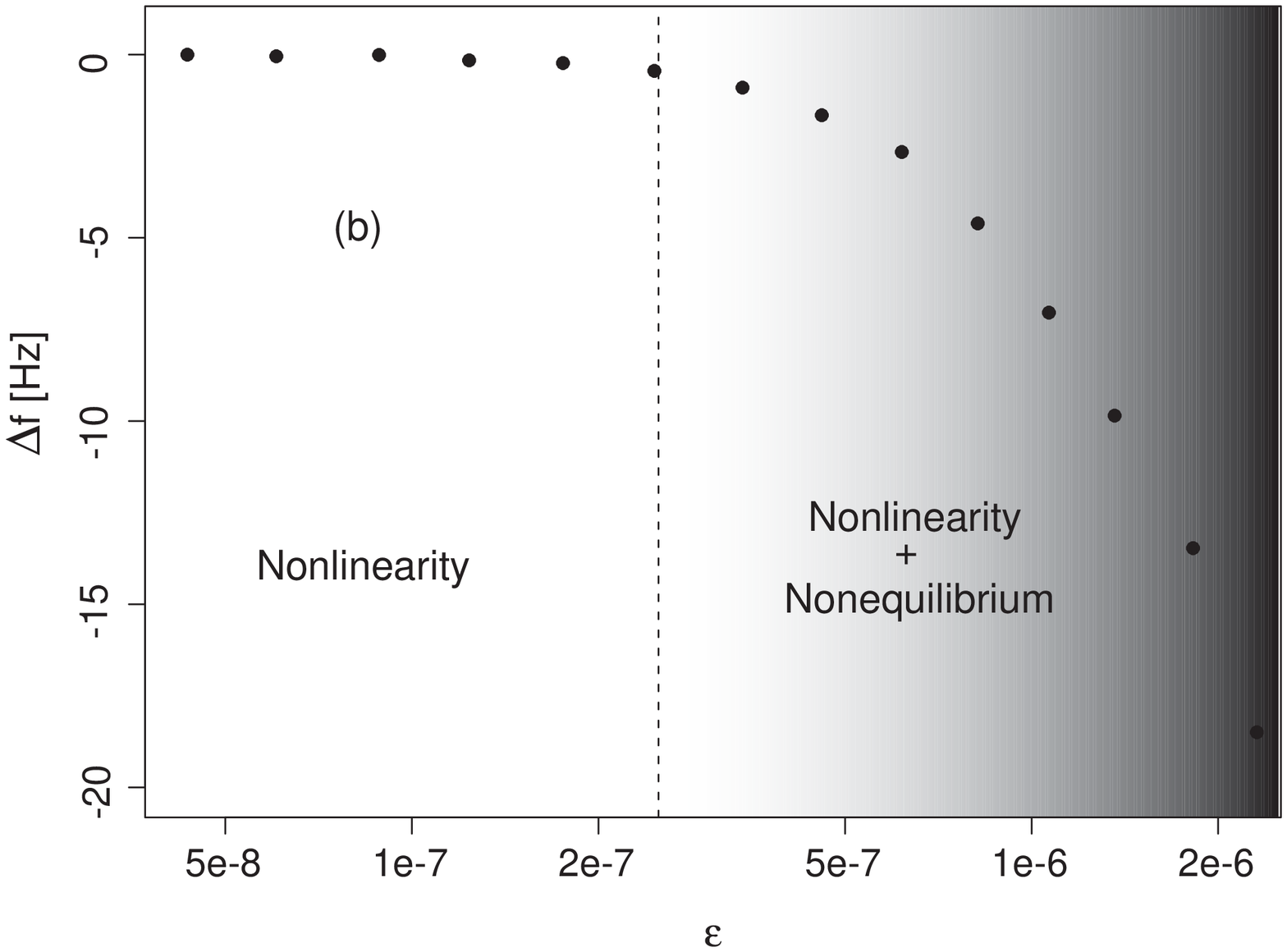} 
\end{center}
\caption{Resonance frequency shift versus strain.  The first regime
where the material displays only an intrinsic reversible nonlinearity
is shown unshaded, and the second regime which combines nonlinear and
nonequilibrium effects is shaded in gray.  The threshold strain for
Berea is $\epsilon_M \simeq $ 5$\cdot 10^{-7}$ (a) and for
Fontainebleau is $\epsilon_M \simeq$ 2$\cdot10^{-7}$ (b).  Since
$\epsilon_M$ is not only a material specific constant but can also
depend on environmental variables, such as temperature and humidity,
we show the regime in which nonlinearity and nonequilibrium are mixed,
not as one solid block, but rather as a region in different shades of
gray.  It is important to note that the data points in the shaded
regions depend on the (temporal) experimental protocol whereas the
data points in the unshaded regions characterize an invariant
behavior.}
\label{fig:2regimes}  
\end{figure} 

Figures~\ref{fig:2regimes} (a) and (b) show our data for the resonance
frequency shifts versus strain for Berea and Fontainebleau samples
respectively.  The first regime, where the material displays only the
intrinsic reversible nonlinearity is shown in the unshaded area,
whereas the regime which combines nonlinear and nonequilibrium
dynamical effects is shaded in gray.  The strain threshold for Berea
is $\epsilon_M \simeq$ 5$\cdot 10^{-7}$ and 2$\cdot 10^{-7}$ for
Fontainebleau under the present experimental conditions.  The data
points in the gray region are history-dependent, and change depending
on the way the experimental protocol is implemented, whereas the data
points in the unshaded region are insensitive to such changes,
provided one begins with the rock in an unconditioned state.  For the
remaining part of the paper we will focus only on the {\em intrinsic
nonlinear regime} which is uncontaminated by conditioning effects and
allows for a simple interpretation of the experimental data.

\subsection{Intrinsic Nonlinearity}

In this section we describe experimental results for strains below
$\epsilon_M$. In this regime, the data are free from memory and
conditioning effects and the samples display a reversible softening of
the resonance frequency with strain. For this reason it is possible to
speak of -- and analyze -- the intrinsic nonlinearity of the
material. As discussed in some detail in the Introduction, the
previous history of resonance measurements and the analysis of the
associated results is somewhat confusing. On the one hand, there are
claims that geomaterials display essentially nonclassical nonlinear
elastic behavior down to very low strains ($10^{-8}$) [{\it Guyer and
Johnson},~1999] with no evidence for a crossover to elastic
behavior. On the other hand, earlier findings [{\it Winkler et
al.},~1979], albeit with generous error bars, are inconsistent with
these claims.

In order to investigate this issue in a systematic and controlled
fashion, we carried out repeatable resonance bar experiments at
strains as low as $10^{-9}$ following the experimental protocols
discussed above; these strains are an order of magnitude lower than
those previously investigated.

\begin{figure}[t]
\begin{center}  
\leavevmode\includegraphics[width=7cm,height=5.5cm]
{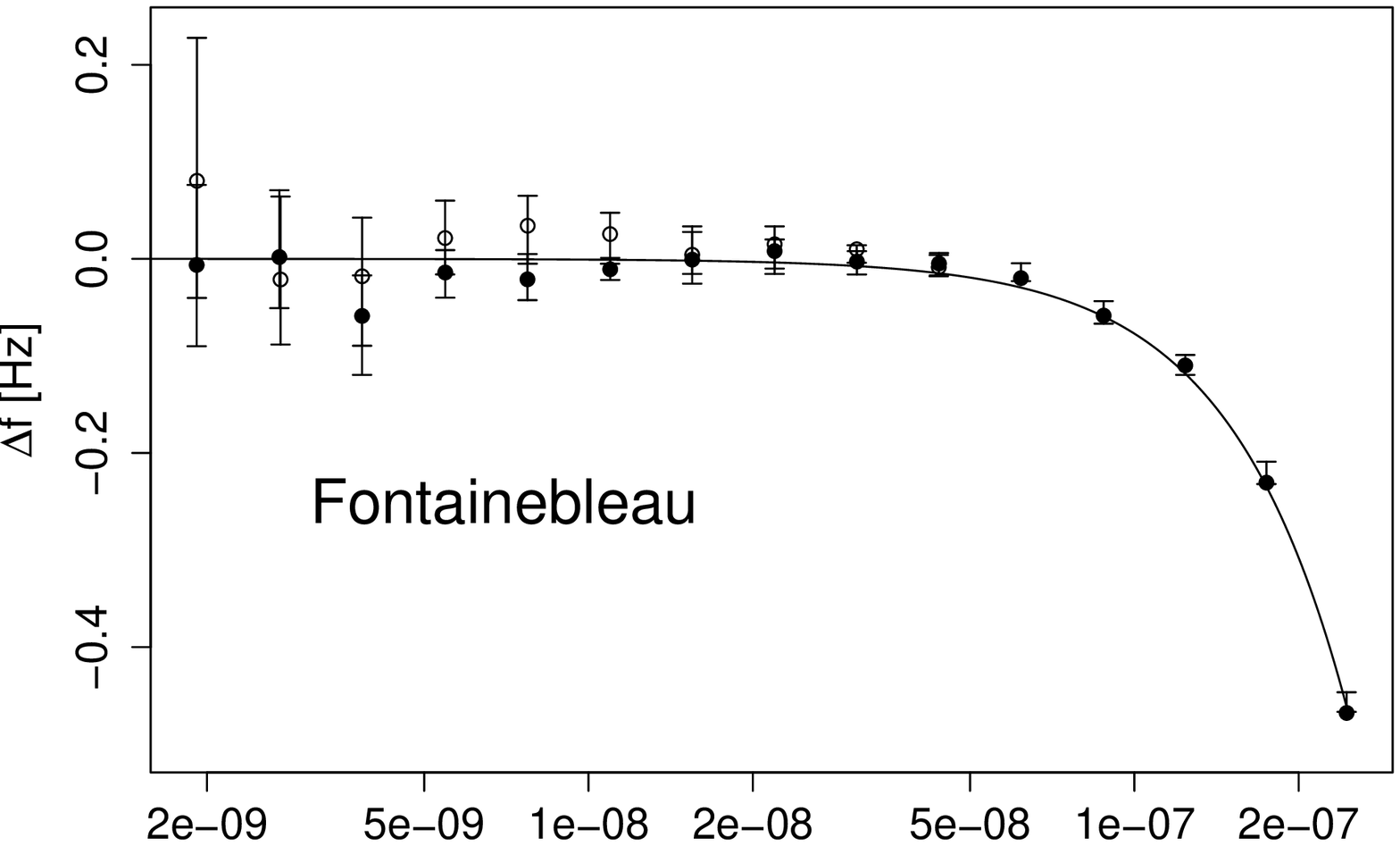} 

\vspace{0.5cm}

\leavevmode\includegraphics[width=7cm,height=5.5cm]
{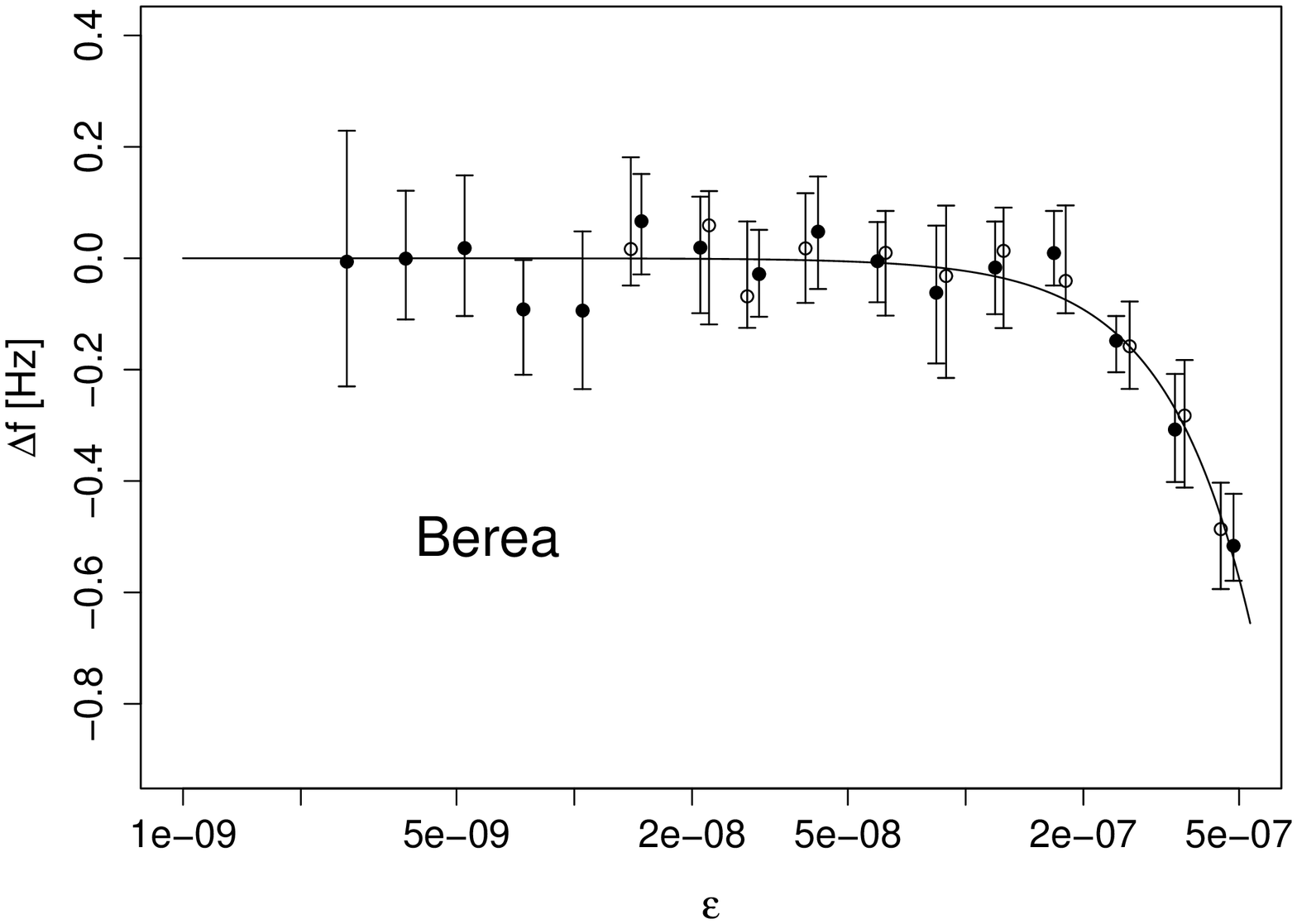}
\end{center}
  \caption{Resonance frequency shift $\Delta f$ as a function of the
  effective strain $\epsilon$ for Fontainebleau and Berea samples for
  $\epsilon < \epsilon_M$.  The solid lines represent predictions of a
  theoretical model incorporating a Duffing nonlinearity,
  Eqn.~(\ref{sig}). Two different sets of data points obtained from
  the same samples are shown to demonstrate the robustness of the
  measurements.  Note the logarithmic scale on the $x$-axis.}
\label{fig:1region} 
\end{figure}

The results for the resonance frequency shift $\Delta f$, $\Delta
f=f_0-{\Omega}/{2\pi}$ where $\Omega$ is the (linear) resonance radian
frequency, as a function of the effective strain $\epsilon$ for
Fontainebleau and Berea sandstone samples are shown in
Figure~\ref{fig:1region}. The measured strain for Fontainebleau ranges
from $2\cdot 10^{-9}$ to $\epsilon_M \simeq 2\cdot 10^{-7}$ and from
$2\cdot 10^{-9}$ to $\epsilon_M \simeq 5\cdot 10^{-7}$ for Berea.  We
observe a resonance frequency shift of 0.45~Hz for Fontainebleau and
0.5~Hz for Berea in the regime below $\epsilon_M$.  The error bars
shown in Figure~\ref{fig:1region} are calculated using the MCMC
analysis as described in Section~\ref{section3}.  The strain error
bars are smaller than the symbols used in the figures.  The error bars
for $\Delta f$ for Berea are larger than the ones for Fontainebleau
because of the smaller $Q$ for the Berea sample: the Berea resonance
curves are much wider, making the peak determination more uncertain.
The solid lines in Figure~\ref{fig:1region} represent the prediction
of a theoretical model with a Duffing nonlinearity described in detail
in Section~\ref{section:model}.

We find that the resonance frequency softens quadratically with
increasing drive amplitude until the strain reaches $\epsilon_M$,
beyond which value conditioning effects also enter.  This behavior can
be fully described by classical nonlinear theory.  At very low
strains, $\epsilon_L\sim 10^{-8}-10^{-7}$ (lower end for
Fontainebleau, upper end for Berea) the samples are effectively in a
linear elastic regime. At these low strains there is no discernible
dependence of the resonance frequency on the strain -- the materials
behave linearly to better than 1 part in $10^4$.  Our results are in
qualitative agreement with previous work by Winkler [{\it Winkler et
al.}, 1979], but in contradiction with other results, {\it Guyer et
al.}~[1999]; {\it Guyer and Johnson}~ [1999], and {\it Smith and
TenCate}~[2000]. We will study this contradiction in detail in
Section~\ref{section:old}.

\subsection{Quality Factor}

Energy loss in solids is mostly characterized by a
frequency-independent loss factor (``solid friction'') in contrast to
liquid friction. Nevertheless, rocks are known to display
characteristics of liquid friction as a function of pore fluid loading
[e.g., {\em Born}, 1941] with an associated dependence of the loss
factor $1/Q$ ($Q$ is also termed the quality factor) on the
frequency. It appears that the unusual nature of wave attenuation in
geosolids remains to be fully studied and understood [Cf. {\em Knopoff
and McDonald}, 1958]. As pointed out by Knopoff and McDonald, a
frequency independent $Q$ cannot be explained by a linear theory of
attenuation, however, it is unlikely that the nonlinearity should be
associated with amplitude since even for very small strains, $Q$
remains finite.

In the present work we do not focus on the dependence of $Q$ on
frequency at small strains, but investigate the dependence on strain
amplitude as an alternative probe of dynamical nonlinearity for
effective strains $\epsilon < \epsilon_M$. We
measure the $Q$ from the amplitude resonance curves directly, using
\begin{equation}\label{qdef}
Q = \frac{\omega_0}{\Gamma}[1+{\cal O}(1/Q^2)]
\label{qdef}
\end{equation}  
where $\omega_0=2\pi f_0$ and $\Gamma$ is the width of the response
curve measured at the points $a_0/\sqrt{2}$ where $a_0$ is the peak
amplitude. This definition of $Q$ is strictly valid only for linear
systems but, as will be discussed further below, at low strains the
amplitude response curves are effectively those of a linear system,
albeit with a peak frequency shift. At leading order, the $Q$ as
defined in (\ref{qdef}) is independent of the nature of the loss
mechanism (solid or liquid friction). 

The loss factor thus depends on two variables, the amplitude response
peak frequency and the width $\Gamma$ of the response curve. We
certainly expect it to change as a function of the strain simply
because $\omega_0$ is a function of the strain amplitude. This is,
however, a very small change, fractionally of order $10^{-4}$. Aside
from this expected variation, what is of more interest is whether
$\Gamma$ is also a function of the strain. 

\begin{figure}
\begin{center}
\leavevmode\includegraphics[width=7cm,height=5.5cm]
{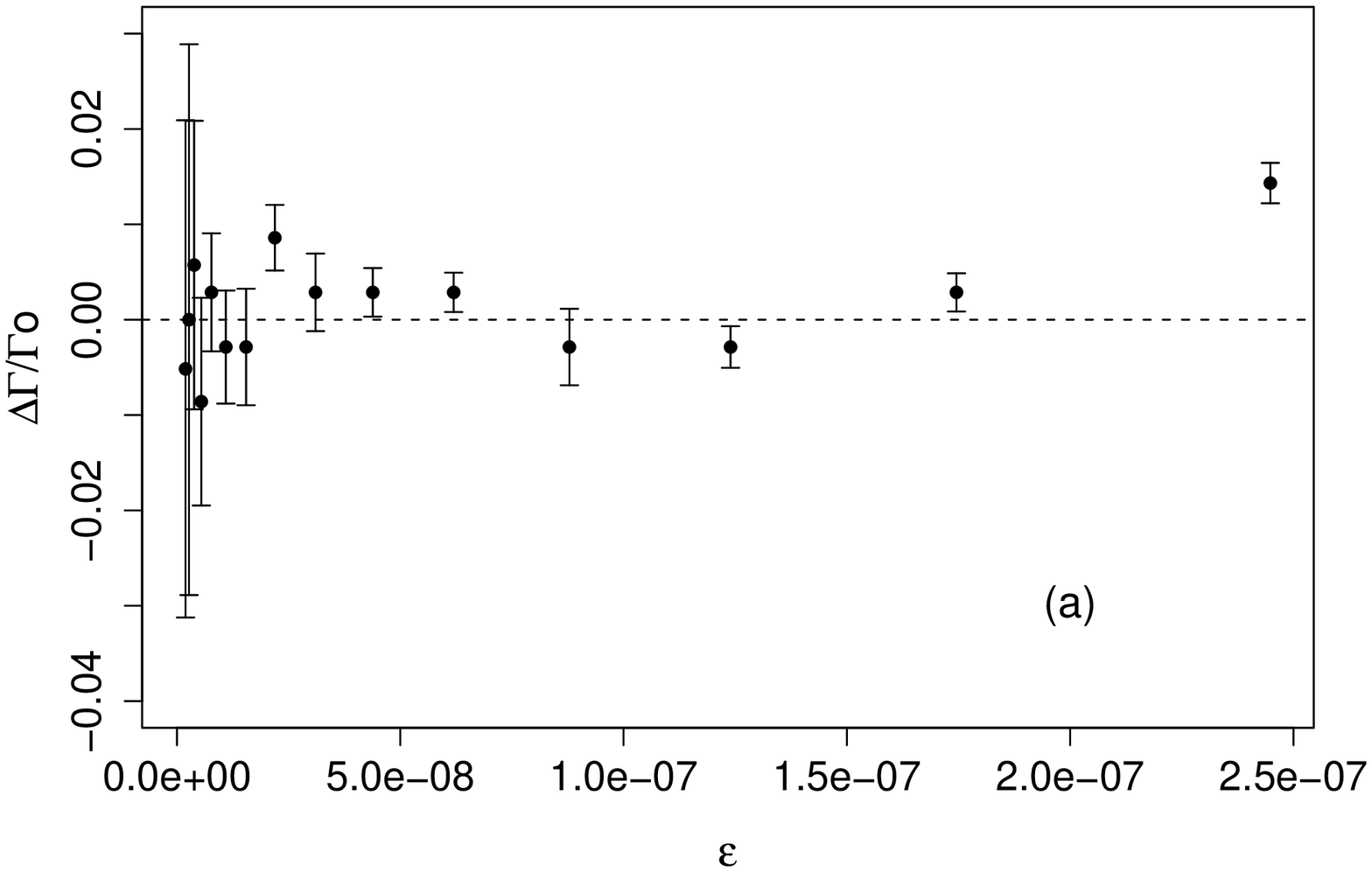} 

\vspace{0.5cm}

\leavevmode\includegraphics[width=7cm,height=5.5cm]
{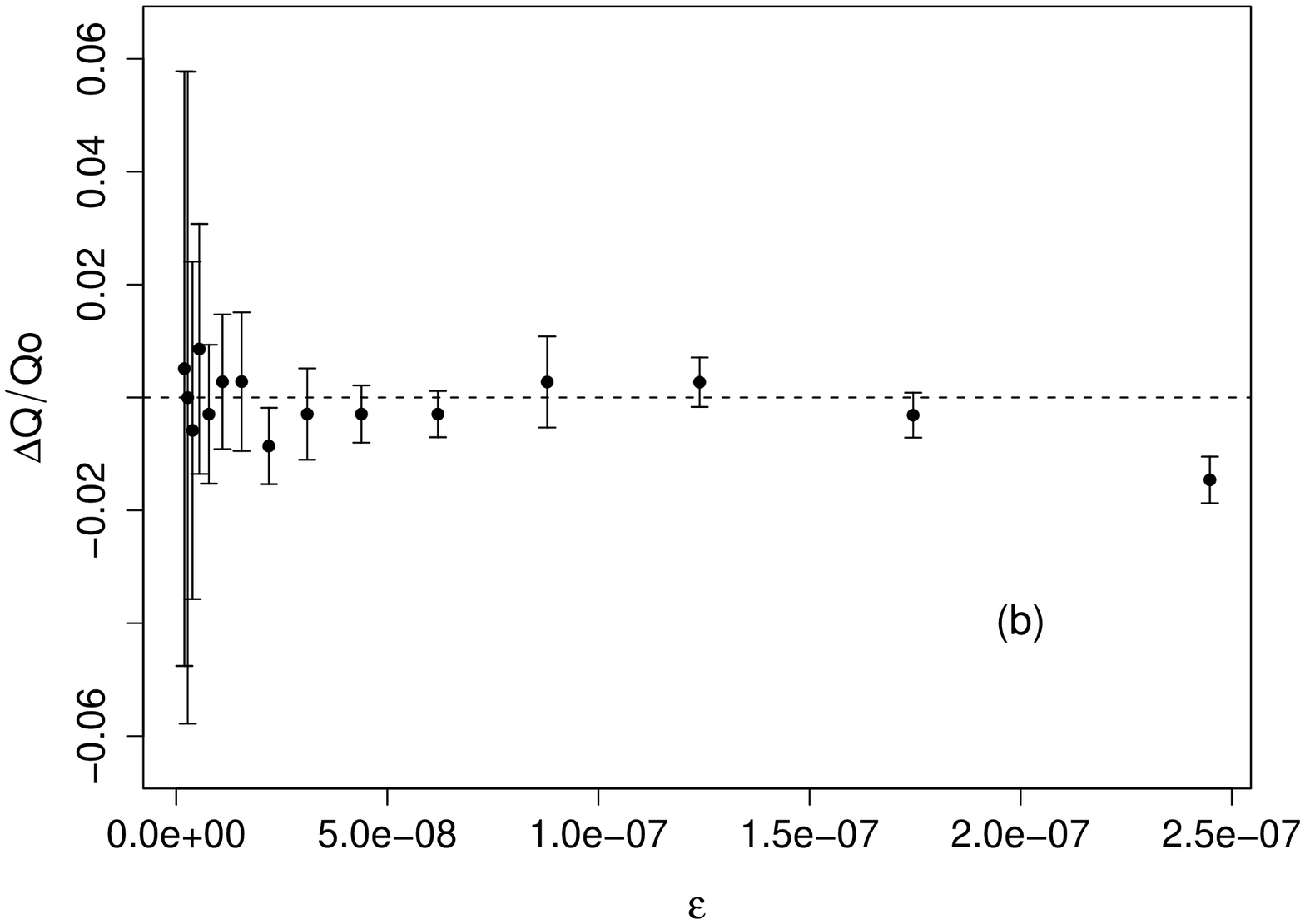}
\end{center}
\caption{Fontainebleau: (a) Variation of the width $\Gamma$ of the
  resonance curve peak. (b) Variation of the quality factor $Q$ with
  strain.} 
\label{fig:Q}
 \end{figure}

In Figure~\ref{fig:Q}(a) we show measurements of the variation in the
relative width $\Delta \Gamma/\Gamma_0$ for the Fountainebleau
sample. As mentioned earlier, we restrict ourselves to the strain
regime below $\epsilon_M$ to prevent contamination of the results by
nonequilibrium effects. The width $\Gamma$ can only be measured to an
accuracy of $\sim 1\%$, the error bars being obtained from MCMC
analysis of the resonance curves. To this accuracy, the results of
Figure~\ref{fig:Q}(a) demonstrate that $\Delta \Gamma/\Gamma_0$ is
essentially constant (except for the single highest strain point) as
is the case for linear systems. This result is also consistent with
the predictions of the Duffing model discussed below in
Section~\ref{section:model}.

The measurement of the relative change in quality factor is shown in
Figure~\ref{fig:Q}(b) and, given the smallness of the frequency peak
shift, simply reflects the behavior of $\Delta \Gamma/\Gamma_0$. We
note that except for the highest strain point, our results are in
agreement with a strain-independent quality factor within the
displayed errors. Our results therefore contradict {\it Guyer et
al.}~[1999] who found a linear dependence of $Q$ on strain amplitude
(over a similar strain range as measured here). To summarize, to the
extent that we have investigated the strain dependence of acoustic
losses ($\epsilon < \epsilon_M$), no unexpected behavior has been
found.

%------------------------------------

\subsection{Stress-Strain Loops and Harmonic Generation}
\label{stressstrainloop}

\begin{figure} 
\begin{center}
\vspace{0.5cm}

\leavevmode\includegraphics[width=7cm,height=5.5cm]
{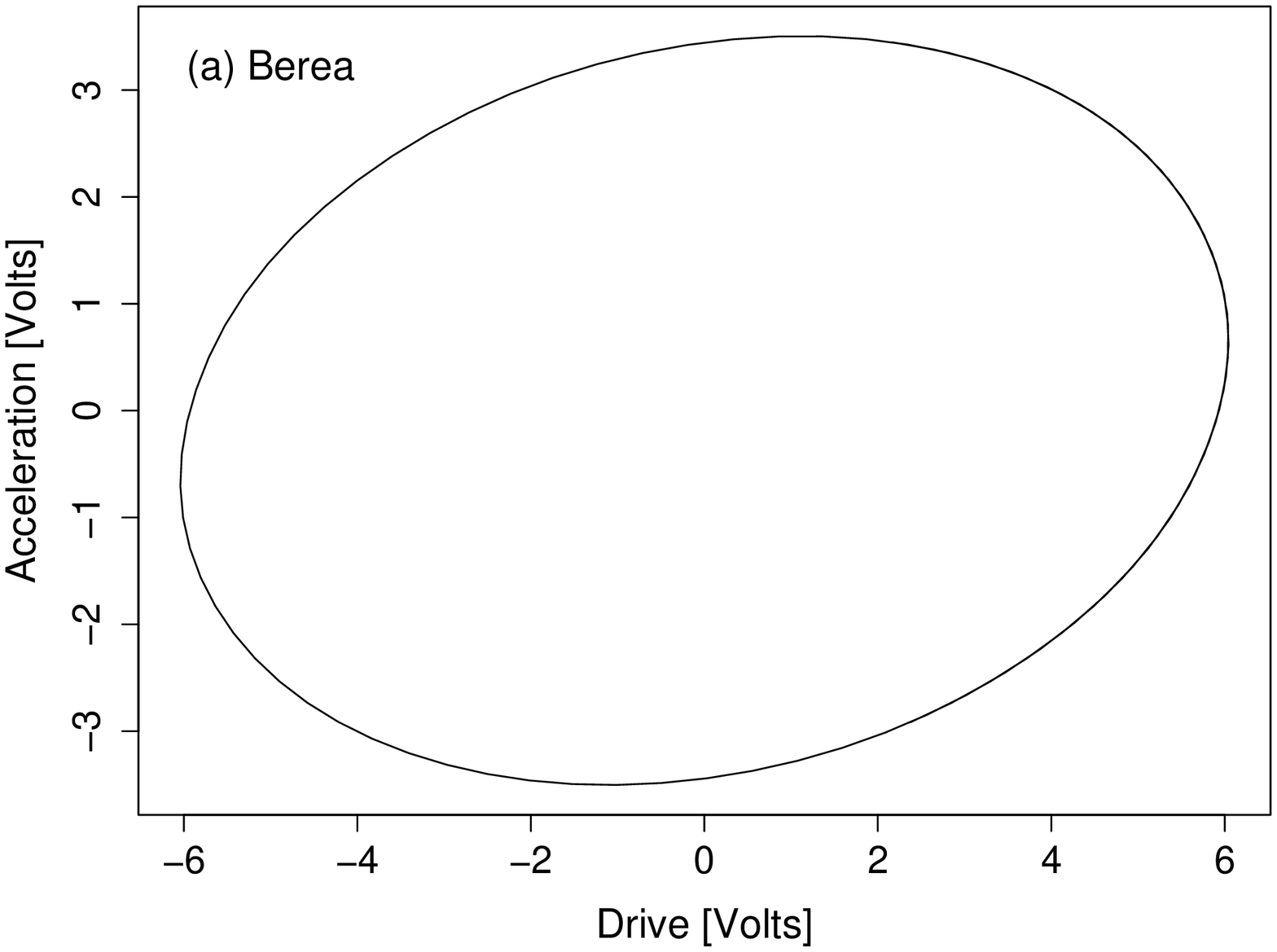}

\vspace{0.5cm}

\leavevmode\includegraphics[width=7cm,height=5.5cm]
{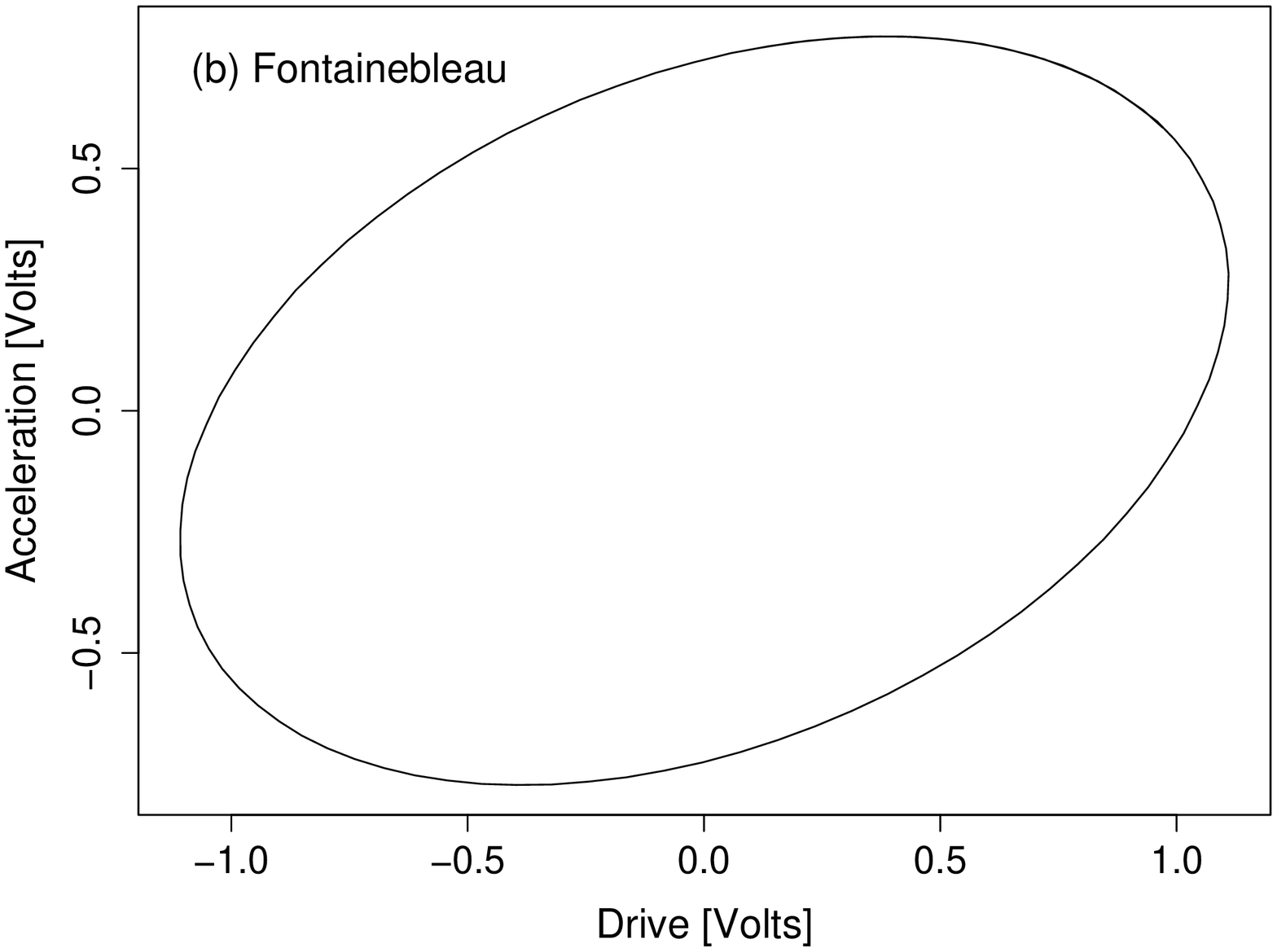} 
\end{center}

\caption{Acceleration versus drive amplitude for the (a) Berea
  and (b) Fontainebleau samples.  The acceleration and the drive
  voltage are proportional to the strain and the stress respectively.
  Berea: strain amplitude $2.5\cdot 10^{-7}$ at a frequency of
  $2754.5$~Hz; Fontainebleau: strain amplitude $10^{-7}$ at a
  frequency of $1154$~Hz. Note the absence of cusps.}
\label{fig:loops}
\end{figure}

At very low strains and at the frequencies of interest here, one would
expect the resonant bar system to be essentially a damped, driven
harmonic oscillator and the hysteresis curve to be an ellipse. This is
in contrast to the situation in (quasi)static hysteresis where
``pointed'' or ``cusped'' loops are observed due to sources of
inelasticity that do not fit in to the simple viscoelastic
model. Whether low strain loops at some point become elliptical was
investigated by {\em MacKavanagh and Stacey} [1974] who came to the
conclusion that this was {\em not} the case at strains $\sim 10^{-5}$
for sandstone and indeed that, ``-- cusped loops extend to
indefinitely small strain amplitudes''. On the other hand, {\em
Brennan and Stacey} [1977] found that for granite and basalt, loops
became elliptical at strain values lower than $10^{-6}$. These
statements were made with data taken at low frequencies, less than
$0.1$ Hz, thus do not directly apply to our experiment unless the
underlying sources of inelasticity continue to be relevant at high
frequencies.

Experimental evidence for cusped stress-strain loops led to the
theoretical description of nonequilibrium dynamics in geomaterials via
PM space models which are based on static-hysteretic building
blocks. In previous work, it has been argued that these models provide
a correct description of the dynamics of rock even at small strains
[{\em McCall and Guyer}, 1994]) and at high frequencies [Cf. {\em
Guyer et al.}, 1999]. Our dynamical experiments allow us to analyze
stress-strain loops at very low strains in the kHz frequency range and
to detect the existence of pointed or cusped loops. As evident in
Figure~\ref{fig:loops} below, the loops are elliptical with no
evidence for cuspy behavior. Thus, we find no evidence to support the
existence of ``nonlinear'' dissipation mechanisms -- as invoked in PM
space models -- at kHz frequencies. In contrast, predictions of the
simple Duffing model introduced in {\it TenCate et al.}~[2004] and
described in detail in Section~\ref{section:model}, are completely
consistent with the data.

Our experimental results are shown in Figure~\ref{fig:loops}.  We plot
the acceleration versus the amplitude of the drive applied to the bar
for both the (a) Berea and (b) Fontainebleau samples. In the case of
Fontainebleau, the strain is $1\cdot 10^{-7}$ at a frequency of
$1154$~Hz while for Berea the strain is $2.5\cdot 10^{-7}$ at a
frequency of $2754.5$~Hz.\footnote{Note that these experiments are
carried out after the original resonance curve measurements were
completed.  Due to different environmental factors, e.g. temperature,
the resonance frequencies of the samples have shifted slightly.} The
acceleration and the drive amplitude are proportional to the strain
and the stress respectively.  The acceleration and the drive voltage
are measured as functions of time and the time series is stored once
steady state was attained. In Figure~\ref{fig:loops}, a piece of the
time series is displayed and the acceleration shifted to obtain it
180$^\circ$ out of phase with the drive voltage.  For both samples,
there is no evidence for cusps in the stress-strain loops.

Another important question is whether the nonlinearity evidenced by
the peak frequency shift can also be detected by searching for
harmonic generation in resonant bar and wave propagation
experiments. The interpretation of results from wave propagation
experiments is somewhat ambiguous [{\em Meegan et al.}, 1993, {\em
TenCate et al.}, 1996] due to experimental complications (e.g.,
reflective losses). However, harmonic detection in (potentially much
cleaner) resonant bar experiments has been previously reported
(Cf. {\em Johnson et al.} [1996]). These authors found substantial
harmonic generation in rock samples -- including Berea and
Fontainebleau -- at strains as low as $10^{-7}$.
 
\begin{figure}
\begin{center}    
\leavevmode\includegraphics[width=7cm,height=5.5cm]
{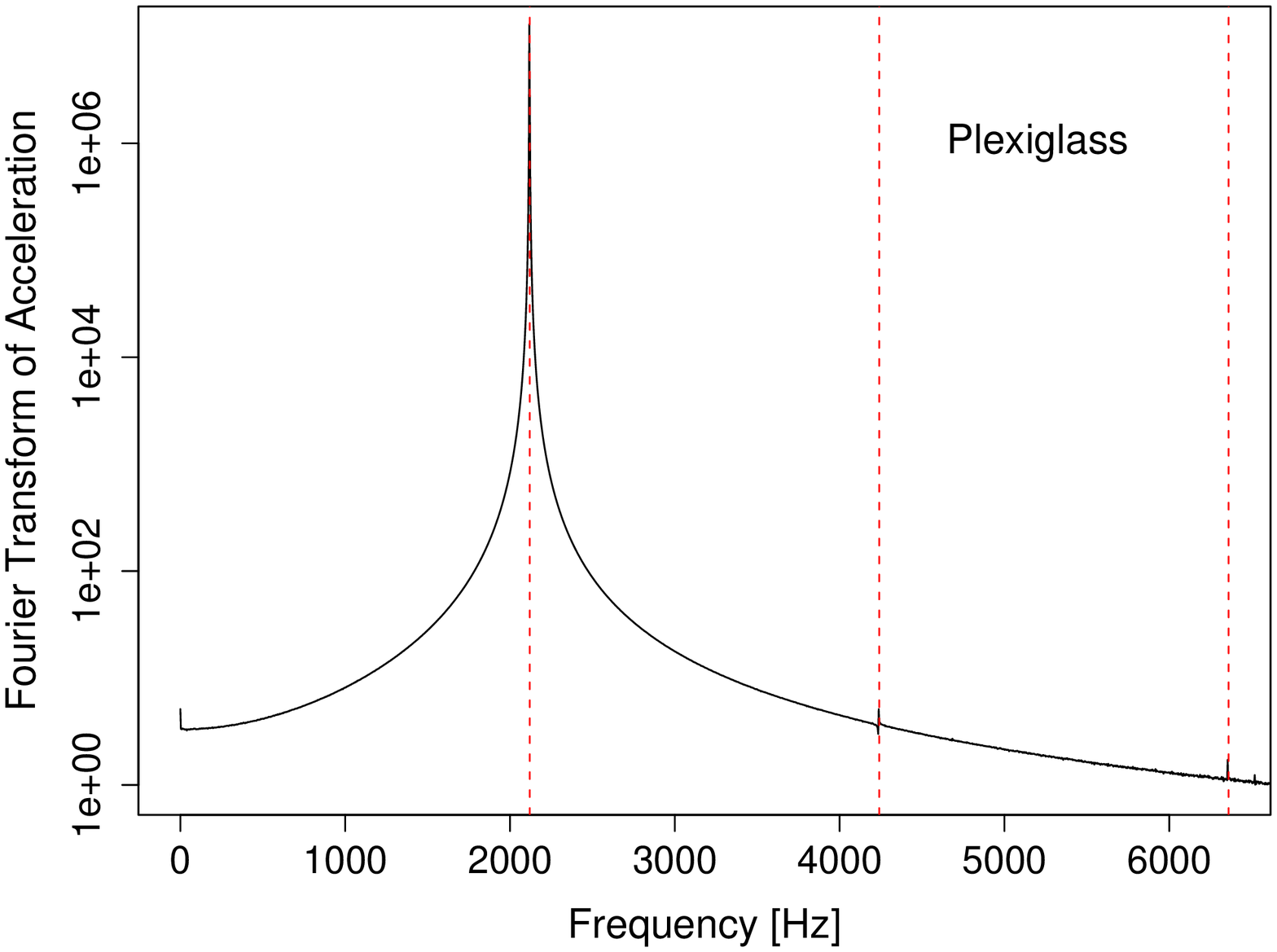} 

\vspace{0.5cm}

\leavevmode\includegraphics[width=7cm,height=5.5cm]
{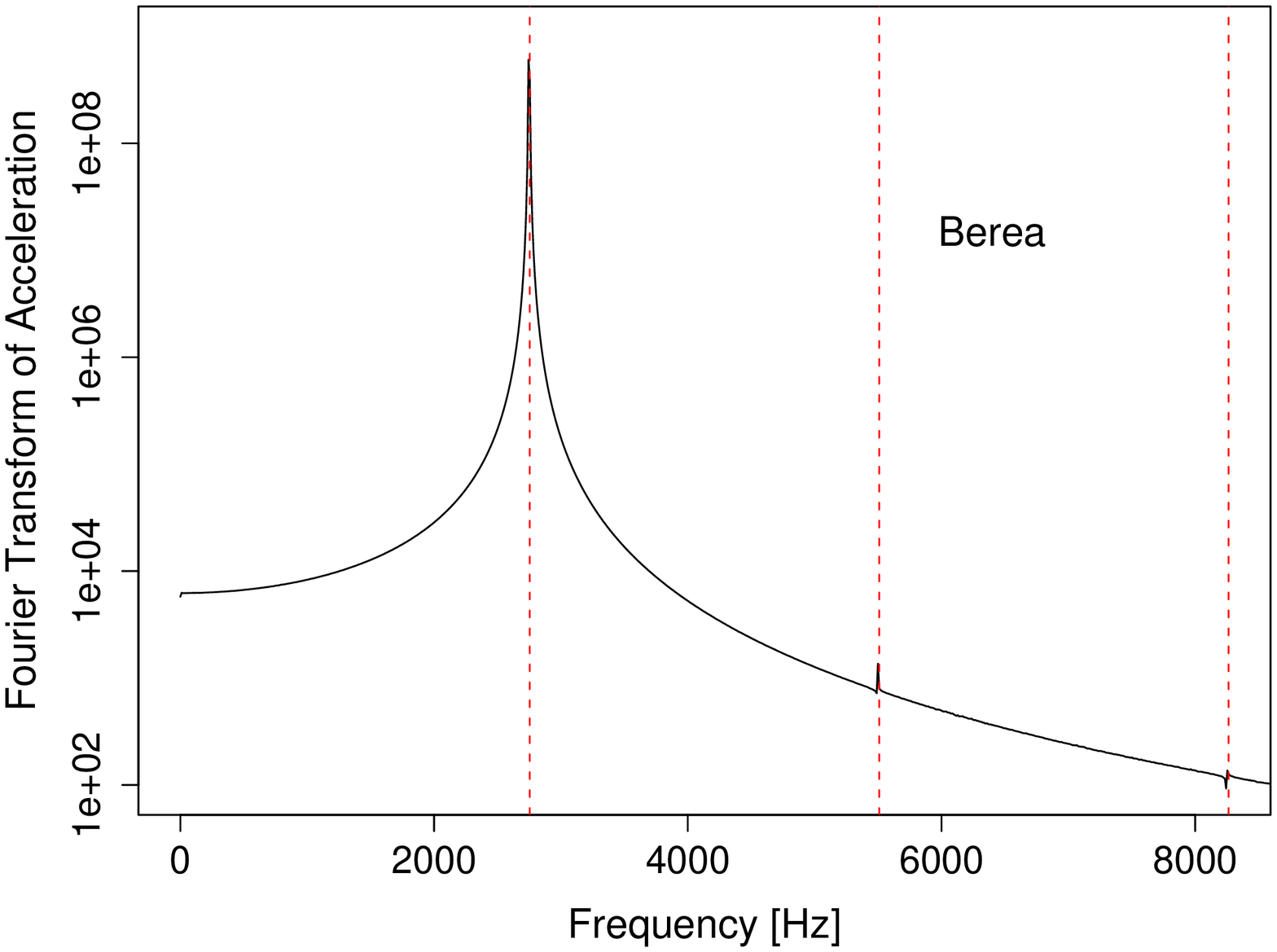} 

\vspace{0.5cm}
 
\leavevmode\includegraphics[width=7cm,height=5.5cm]
{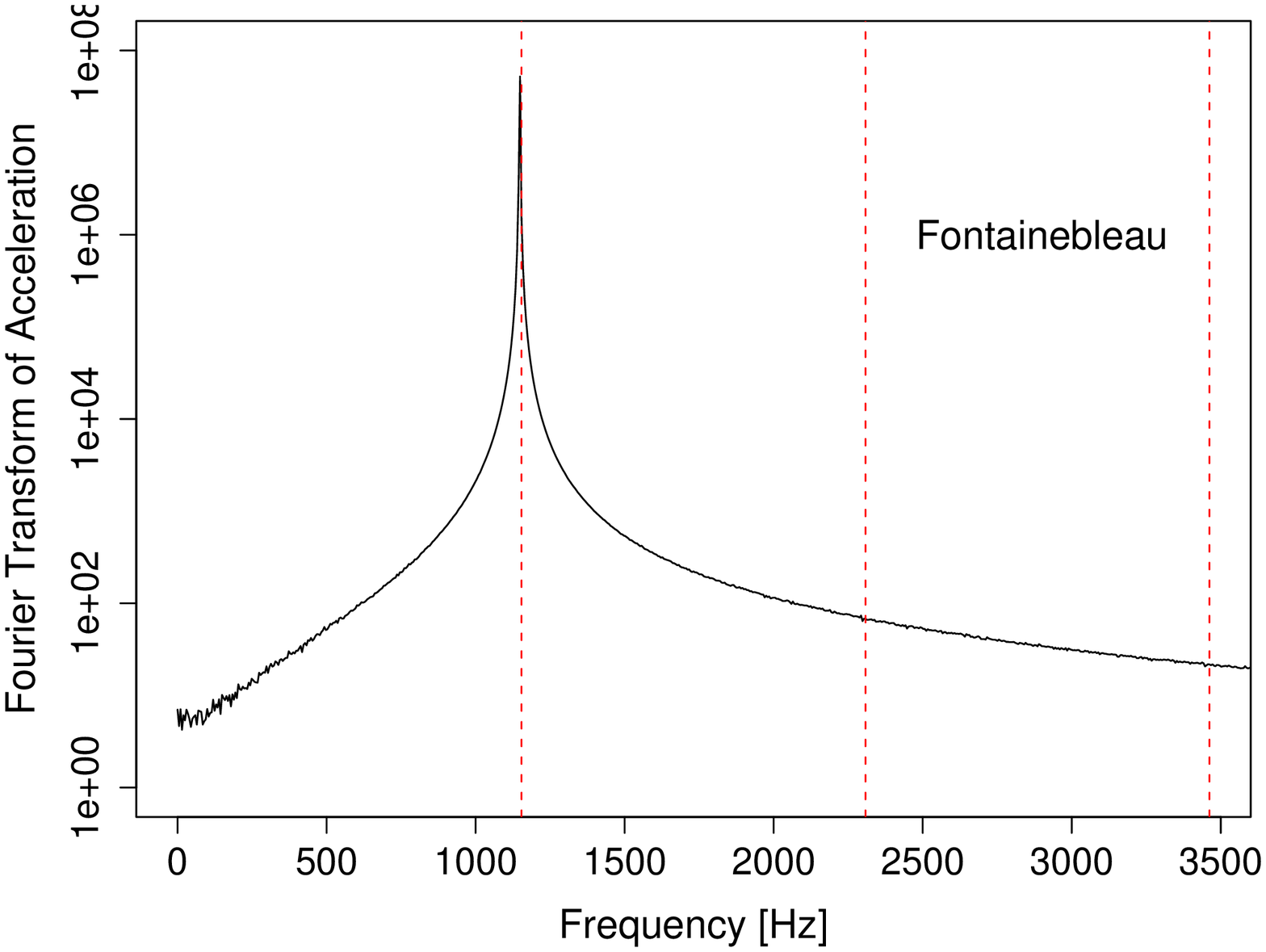}  
\end{center}
\caption{Fourier transform of the acceleration taken at the resonance
frequency for Acrylic, Berea and Fontainebleau (semilog plot).
Acrylic: nominal strain of $2.6 \cdot 10^{-6}$ at frequency $2120$Hz;
Berea, $2.5\cdot 10^{-7}$ at $2754.5$~Hz; Fontainebleau, $10^{-7}$ at
frequency $1154$~Hz. The dashed lines show the positions of the first,
second and third harmonics. Harmonic generation is not detected. The
two spikes which occur in Plexiglass and Berea are due to the residual
nonlinearity of the experimental apparatus.}
\label{fig:harmonics} 
\end{figure}

In this paper, we present our results in a search for harmonics at
strains $\epsilon < \epsilon_M$. Figure~\ref{fig:harmonics} shows
spectral measurements for a linear material (acrylic) and the two rock
samples. The dashed lines indicate where the first, second, and third
harmonics of the fundamental are expected to appear (these are not the
higher Pochhammer modes). In all three cases we observe no evidence
for the existence of higher order harmonics. The two small spikes
which occur in the data for Plexiglass (acrylic) and Berea are due to
the residual nonlinearity of the experimental apparatus.

\section{The Model}\label{section:model}

In this section we introduce a simple phenomenological model which
describes the nonlinear behavior of the rock samples under
consideration.  This model does not include a treatment of memory and
nonequilibrium effects and is therefore not meant to apply in the
regime where these effects become important, i.e. for strains greater
than $\epsilon_M$.  A more complex model which applies also to the
higher strain regimes will be described elsewhere.  As shown by us
previously ({\it TenCate et al.},~[2004]), a quartic (Duffing)
potential nonlinearity augmenting a damped harmonic oscillator yields
results that accurately describe the data in the low strain regime.
This model predicts a quadratic softening of the resonance frequency
as a function of drive amplitude, as expected from the theory of
classical nonlinear elasticity.

The equation of motion for the displacement is taken to be:
\begin{equation}
\ddot u+ \Omega^2 u + 2\mu\dot u+ \gamma u^3 = F \sin(\omega t),
\label{EOM}
\end{equation}
where $\gamma < 0$ leads to a softening nonlinearity as observed in
the experiment (e.g., Figure~\ref{fig:RC}). The driving force on the
right hand side represents the drive applied to the rods in the
experiment. The frequency $\Omega$ is the (unshifted) harmonic
oscillator frequency (for $\gamma,\mu=0$) and $\mu$ is the linear
damping coefficient. In the following we briefly discuss a convenient
analytic approximation for the solution of Eqn.~(\ref{EOM}).

\subsection{Multiscale Analysis}
Since the displacement $u$ is small we can solve the equation of
motion (\ref{EOM}) analytically and predict the softening of the
frequency with the drive amplitude.  We employ multiscale perturbation
theory to obtain a useful closed-form solution to Eqn.~(\ref{EOM}).
In the following we describe how this approach works and how to
extract model parameters from experimental data.  [For a complete
derivation of multiscale perturbation theory see~{\it Nayfeh},~1981.]
While one can of course solve Eqn.~(\ref{EOM}) numerically, the
analytic approach yields simple formulae which provide much better
physical intuition.

A naive approach to solving Eqn.~(\ref{EOM}) would be a
straightforward expansion of the displacement in the form
\begin{equation}
u(t,\alpha) = u_0(t)+\alpha u_1(t)+\cdots. \label{expan}
\end{equation}
This ansatz is justified for small displacements. Inserting the
expansion of $u$ in the equation of motion and keeping only terms of
${\cal O}(\alpha)$ leads to two differential equations for $u_0$ and
$u_1$: 
\begin{eqnarray}
&&\ddot u_0+\Omega^2 u_0 = F \sin(\omega t),\label{u0}\\ 
&&\ddot u_1+\Omega^2 u_1 = -2\mu\dot u_0-\gamma u_0^3,\label{u1}
\end{eqnarray}
which are simply harmonic oscillators with an inhomogeneity on the
right hand side.  The equation for $u_0$ (\ref{u0}) can be solved
immediately and the solution inserted into the right hand side of the
equation of motion for $u_1$ (\ref{u1}) specifying the inhomogeneity
for $u_1$ completely.  The solution for $u_1$ can now be determined
and a perturbative solution for $u$ itself can be obtained by
inserting $u_0$ and $u_1$ into Eqn.~(\ref{expan}).  A detailed
analysis of this solution for $u(t)$ leads to the following result:
for specific values of $\omega$ resonances occur, the case
$\omega\sim\Omega$ leading to a primary resonance causing the solution
for $u$ to diverge.  To determine a solution for Eqn.~(\ref{EOM}) free
from this problem, the method of multiple scales can be used [{\it
Nayfeh},~1981]. The idea is the following: besides assuming that the
displacement is small, we also assume that the nonlinearity is small.
In addition we assume that the excitation, the damping, and the
nonlinearity are all of the same order in $\alpha$. This leads to a
modified equation of motion for $u$:
\begin{equation}
\ddot u +\Omega^2 u + 2\alpha\mu\dot u+ \alpha\gamma u^3 =
 \alpha F \sin(\omega t).
\end{equation}
Further we introduce two time scales, a slow scale
$T_1=\alpha t$ and a fast time scale $T_0=t$ which leads to a
transformation of the derivatives of the form 
\begin{eqnarray}
\frac{d}{dt} &=& D_0 +\alpha D_1,\\
\frac{d^2}{dt^2} &=& D_0^2+2\alpha D_0D_1+\cdots,
\end{eqnarray}
with $ D_i=\partial/\partial T_i$. Expanding $u$ in the form
\begin{equation}
u=u_0(T_0,T_1)+\alpha u_1 (T_0,T_1)
\end{equation}
and keeping again only terms of order $\alpha$ leads to the following
set of differential equation for $u_0$ and $u_1$: 
\begin{eqnarray}
D_0^2 u_0+ \Omega^2u_0 &=& 0,\\
D_0^2 u_1+ \Omega^2u_1 &=& -2 D_0 D_1 u_0 \label{u1new}\\
&&- 2\mu D_0 u_0
-\gamma u_0^3+F\sin (\omega T_0).\nonumber
\end{eqnarray}
The difference with the previous naive expansion becomes clear
immediately: While earlier the driving force was part of the
differential equation for $u_0$, it is now part of the inhomogeneity
of $u_1$. A general solution for $u_0$ is given by
\begin{equation}
u_0= A(T_1)e^{iT_0}+\bar A(T_1) e^{-iT_0}.
\label{u0exp}
\end{equation}
Inserting Eqn.~(\ref{u0exp}) into the differential equation for
$u_1$ (\ref{u1new}) yields
\begin{eqnarray}
D_0^2u_1+\Omega^2 u_1&= & -(2iA'\Omega+2i\mu A\Omega+3A^2\bar A
\gamma)e^{i\Omega T_0}\nonumber  \\   
&& -A^3\gamma e^{3i\Omega T_0} +\frac 1 2 F e^{i\omega T_0} + 
\mbox{c.c.}
\label{u1final}
\end{eqnarray}
Since we are only interested in the case $\omega\sim\Omega$, i.e.,
driving near to the resonance frequency we introduce a detuning
parameter
\begin{equation}
\label{detune}
\omega=\Omega+\alpha\sigma~~\Rightarrow ~~\omega T_0=\Omega 
T_0+\sigma T_1.
\end{equation}
Inserting this expression into the differential equation
(\ref{u1final}), expressing $A$ in the polar form
$A=1/2a\exp{i\beta}$, defining a new parameter $\phi = \sigma
T_1-\Omega\beta$ and $\phi ' = \sigma-\Omega\beta '$, and eliminating
the secular terms from the resulting equation, we arrive at the
following solution for $u(t)$:
\begin{eqnarray}
u&=&a \cos (\omega t -\phi)+{\cal O}(\alpha),\\
a'&=&-a \mu +\frac 1 2 \frac{F}{\Omega}\sin{\phi},\label{aprime}\\
\label{phiprime}a\phi'&=&a\sigma-\frac 3 8\frac{\gamma a^3}{\Omega}
+\frac{1}{2}\frac{F}{\Omega}\cos (\phi).
\end{eqnarray}
After a sufficiently long time, $a$ and $\phi$ will reach a
steady-state hence their derivatives will vanish and the left hand
sides of Eqns.~(\ref{aprime}) and (\ref{phiprime}) will be
zero. Squaring the equations and adding them leads to the so-called
frequency-response equation
\begin{equation}\label{freq}
\Omega^2\mu^2a^2 + a^2\left(\sigma\Omega
-\frac 3 8 a^2\gamma \right)^2=\frac 1 4 F^2.
\end{equation} 
This equation can be solved with respect to $\sigma$
\begin{equation}
\sigma=\frac{3}{8}a^2\frac{\gamma}{\Omega}
\pm\frac{1}{2a\Omega}\sqrt{F^2-4\mu^2a^2\Omega^2}.
\end{equation}
As $\sigma$ has to be real, the maximum value for $a$ (which we label
$a_0$) and therefore the peak of the response curve can be immediately
determined:
\begin{equation}
F^2=4\mu^2 a_0^2\Omega_0^2 ~~\Rightarrow~~a_0=\frac{F}{2\mu\Omega},
\end{equation}
and therefore
\begin{equation}\label{sig}
\sigma_0=\frac{3F^2\gamma}{32\mu^2\Omega^3}.
\end{equation}
Thus, the model predicts a quadratic softening of the frequency with
the drive amplitude $F$. The model also predicts the invariance of the
resonance curve width $\Gamma$ for any strain.  Solving 
Eqn.~(\ref{freq}) for $\sigma$ and substituting $a=a_0/\sqrt{2}$ we
obtain
\begin{equation}\label{gamma}
\Gamma=2 \mu. 
\end{equation}
Note that the approximation ignores corrections of
${{\cal O}(1/Q^2)}$. These are numerically small on the scale of the
experimental errors. At this leading order of the approximation, the
effect of the nonlinearity is simply to produce an effective harmonic
oscillator response, with a frequency shift and peak height dependent
on the drive amplitude.

\subsection{Constraints on the Model Parameters from the Experimental Data}
The Duffing model predicts an invariant resonance curve width
$\Gamma$, therefore we first measure this quantity from the
experimental resonance curves. Consistent with the above expectation,
we find that $\Gamma$ is constant within $1\%$ for both samples over
the applicable strain range; using relation (\ref{gamma}) we then
immediately determine the damping coefficients $\mu = 27.5$~s$^{-1}$
for the Fontainebleau and $\mu=131.6$~s$^{-1}$ Berea sample,
respectively.  Using the definition of $\sigma_0=2\pi f_0 - \Omega$
and the relation $F=2\mu \Omega L\epsilon/\pi$ we can rewrite
Eqn.~(\ref{sig}) in terms of the effective strain $\epsilon$ and the
resonance frequency $f_0$ as:
\begin{equation}
f_0 =\frac{3 L^2 \gamma}{16\pi^3 \Omega}\epsilon^2 + 
\frac{\Omega}{2\pi}.
\end{equation}
The linear resonance frequency $\Omega$ and the nonlinearity parameter
$\gamma$ now follow by fitting the experimental data for $f_0$ as a
function of the effective strain using the previous equation.  We
obtain the following values: the nonlinearity parameter,
$\gamma=-7.6\cdot 10^{19}$~m$^{-2}$s$^{-2}$ for the Fontainebleau
sample, and $\gamma=-5.3\cdot 10^{19}$~m$^{-2}$s$^{-2}$ for the Berea
sample, whereas the corresponding linear resonance frequencies are
$7262.8$~rad/s and $17375.7$~rad/s.  
  
\subsection{Comparison of the Experimental Results with the Model}

After determining model parameters as above, we compare the Duffing
model predictions with the experimental results described in
Section~\ref{expres}.

\begin{figure}
\begin{center}
\leavevmode\includegraphics[width=7cm,height=5.5cm]
{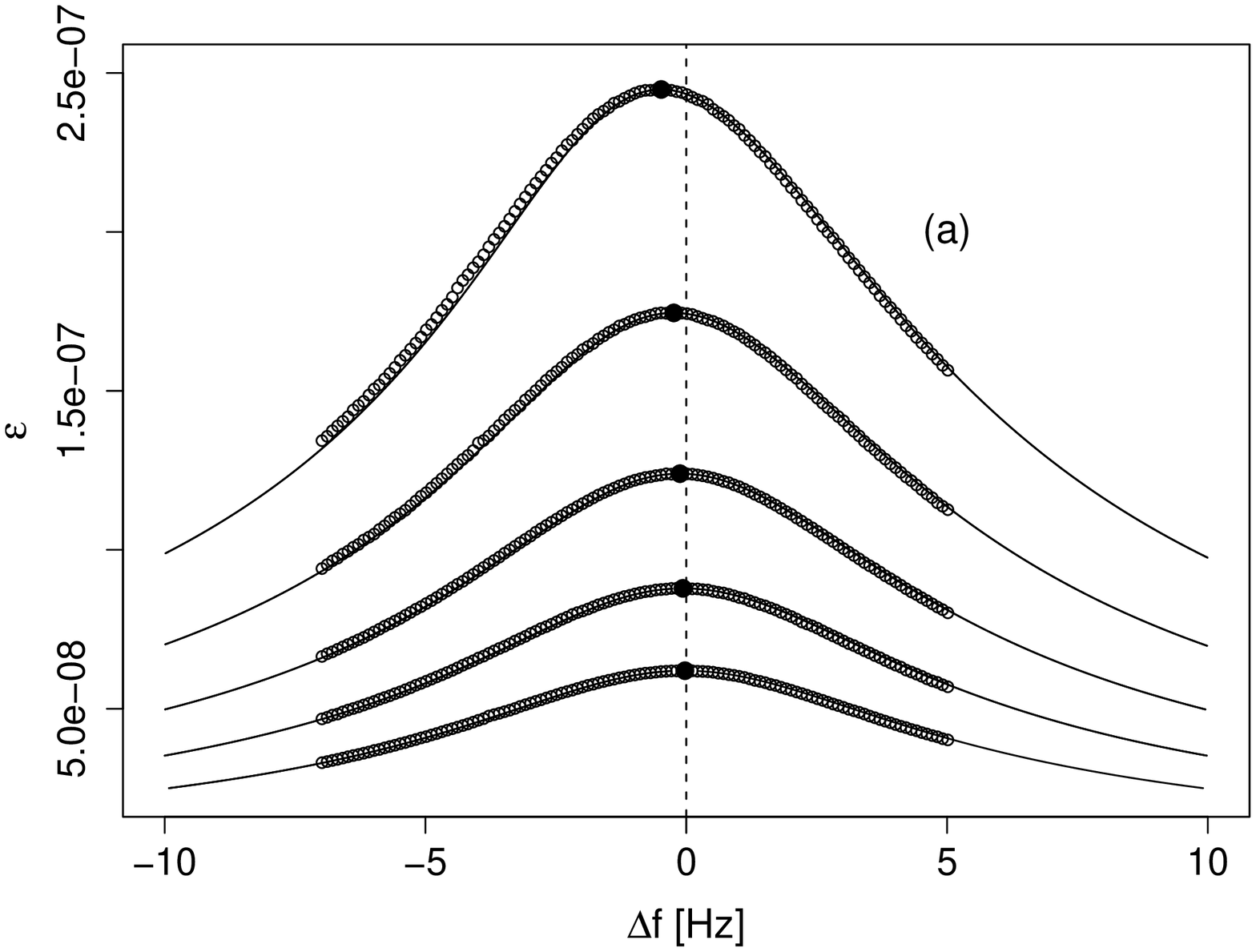} 

\vspace{0.5cm}

\leavevmode\includegraphics[width=7cm,height=5.5cm]
{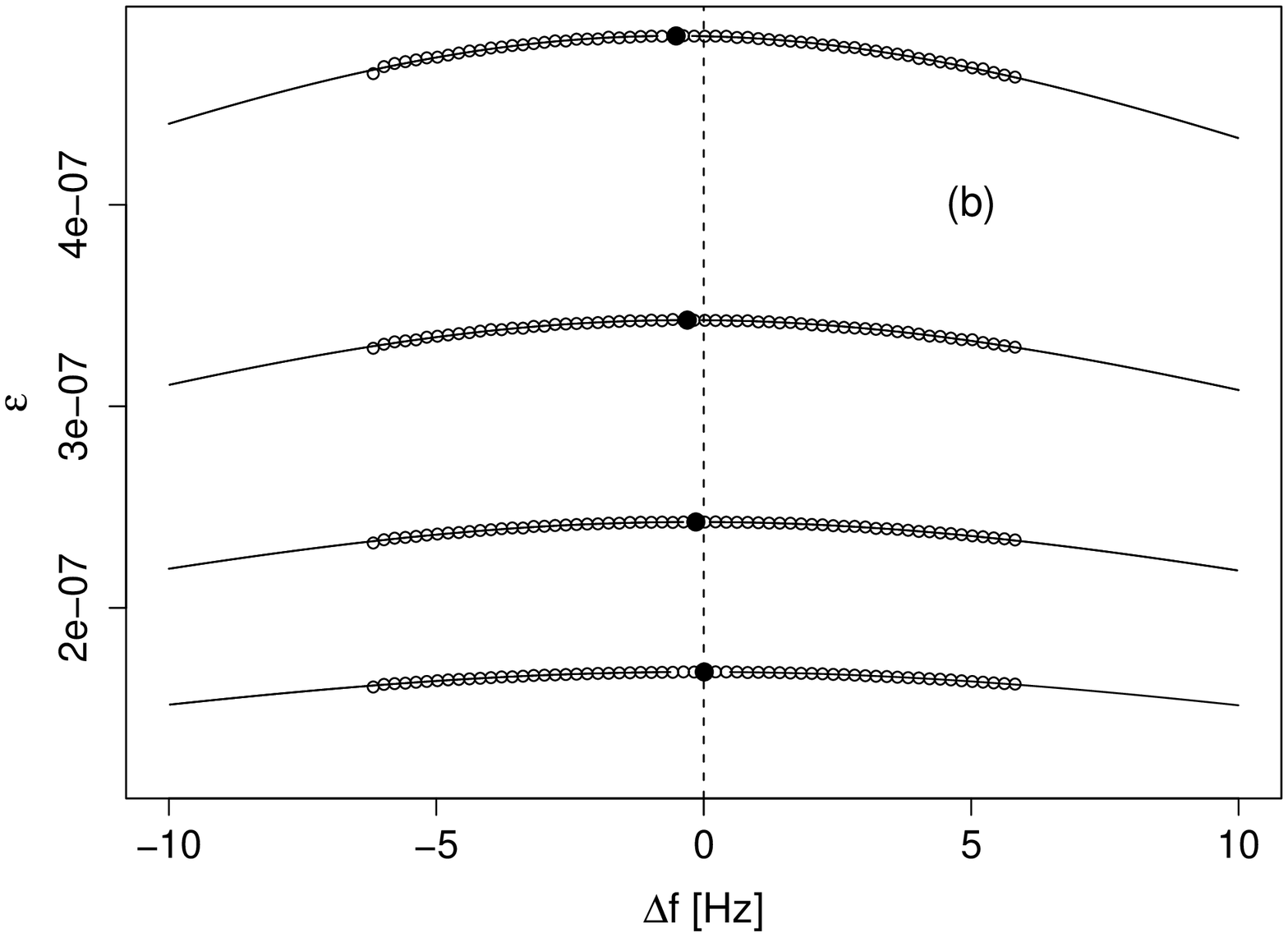} 

\vspace{0.5cm}

\leavevmode\includegraphics[width=7cm,height=5.5cm]
{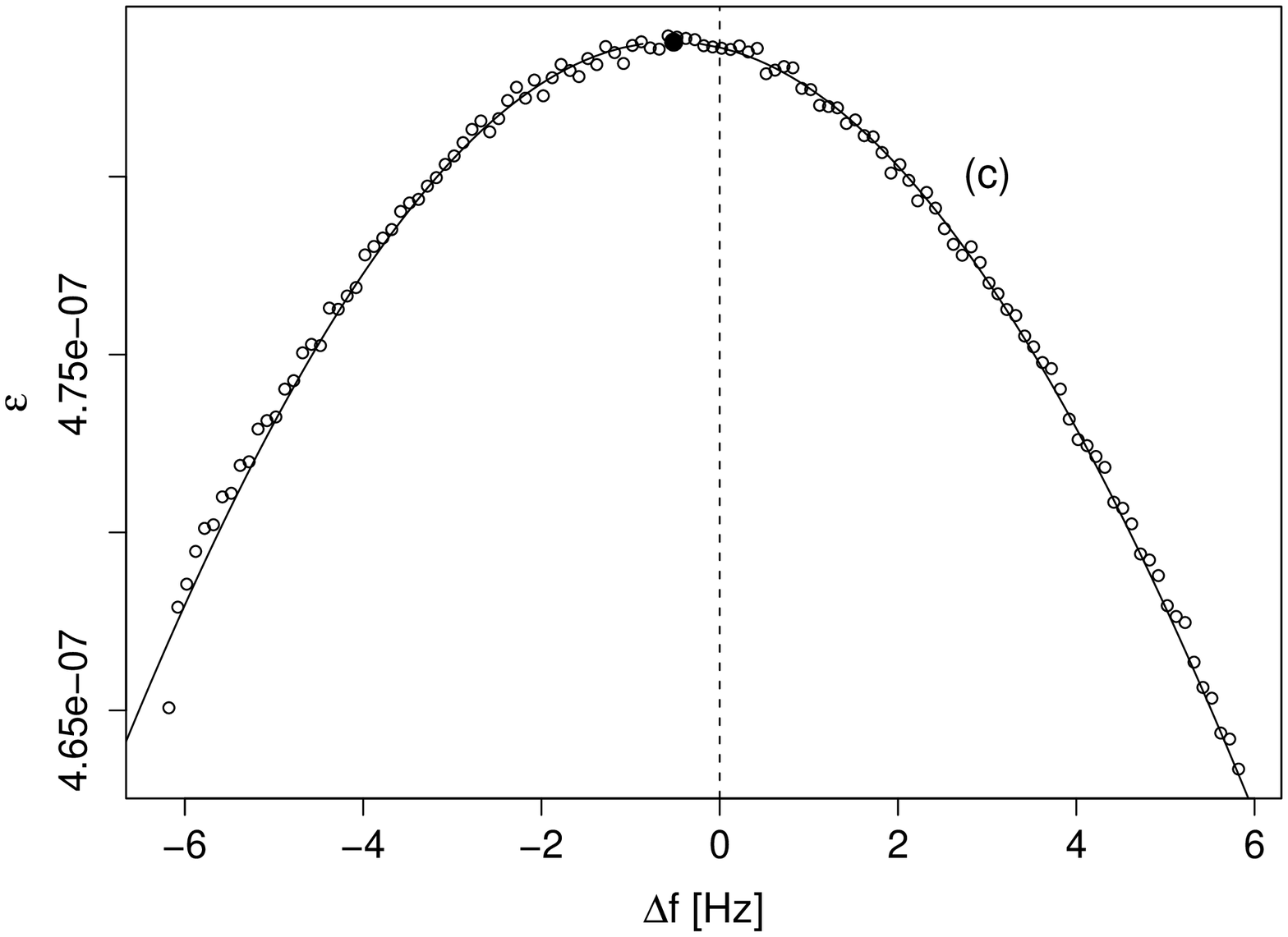} 
\end{center}
\caption{Average strain amplitude $\epsilon$ as a function of drive
frequency for Fontainebleau (a) and Berea (b) and (c).  The reference
center frequency is 1155.98~Hz for Fontainebleau and 2765.179~Hz for
Berea.  The open circles are the experimental data; the filled circles
mark the peak positions.  The solid lines are theoretical predictions
from Eqn.~(\ref{freq}).  Figure~(c) shows in detail the resonance
curve at the highest strain for Berea.}
\label{fig:model} 
\end{figure}

We begin by investigating the predictions for the resonance curves
themselves, as given in Eqn.~(\ref{freq}).  In Figure~\ref{fig:model}
we show the results from the experiments as circles and the results
from the Duffing model as solid lines for (a) Fontainebleau and (b)
Berea, where (c) shows a single Berea resonance curve on a smaller
range in $\Delta f$ to demonstrate more clearly how well the model
works. In addition, it was shown earlier (Figure~\ref{fig:1region})
how the resonance frequency shifts as a function of strain for
Fontainebleau and Berea from both the experiment and the model.
Figure~\ref{fig:1region} and Figure~\ref{fig:model} clearly
demonstrate the excellent agreement between the experimental data and
the model predictions.

\begin{figure}
\begin{center}
\leavevmode\includegraphics[width=7cm,height=5.5cm]
{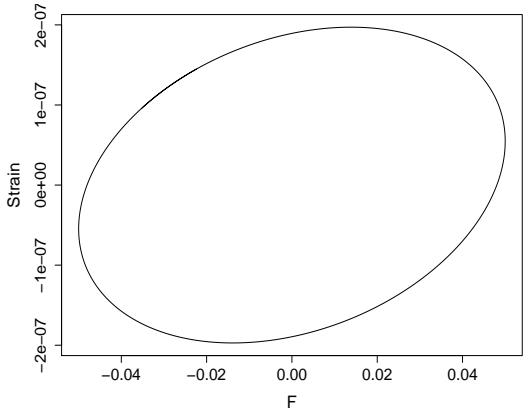}
\end{center}
\caption{Hysteresis loop as predicted by the Duffing model using Berea
  parameters, strain $2.7 \cdot 10^{-7}$, frequency = 2765.3~Hz. }  
\label{fig:loop_model} 
\end{figure}

\begin{figure}
\begin{center}
\leavevmode\includegraphics[width=7cm,height=5.5cm]
{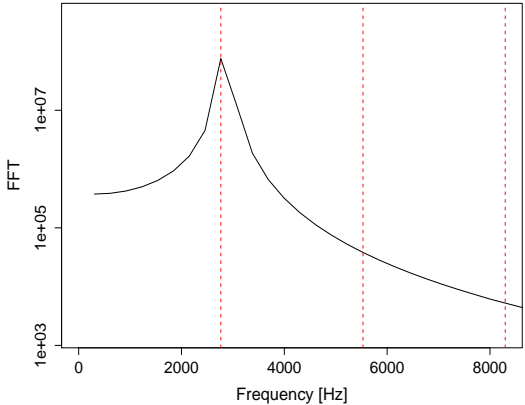}
\end{center}
\caption{Spectral response from the Duffing model using Berea
  parameters, strain $2.7 \cdot 10^{-7}$,  frequency = 2765.3~Hz. }  
\label{fig:harm_model} 
\end{figure}

In Figure~\ref{fig:loop_model} we show the stress-strain loop obtained
from the Duffing model: no cusps are present in agreement with the
experimental results.  Moreover our model indicates that the response
of the bar to the external drive is dominated by the fundamental
mode and there is no excitation due to mode-coupling of any higher
harmonics as shown in Figure~\ref{fig:harm_model}.  This prediction is
in contradiction with previous work [{\it Johnson et al.},~1996]
where it was claimed that the absence of frequency softening is not
sufficient to rule out nonlinearity in rocks as harmonic generation
may exist even in the absence of a discernible frequency shift.  Our
model predictions are again in very good agreement with the 
experimental results.

%%%%%%%%%%----------------
\section{Comparison with Previous Results}\label{section:old}

%------------------------------------

As already discussed in the Introduction, experiments similar to the
one described in this paper have been carried out in the past with
somewhat confusing results.  Some of them, e.g., those of {\em Winkler
et al.}~[1979], are in qualitative agreement with our findings though
with less control over errors, while other papers claim quite
different results.  Among this second set of papers, two papers are
experimentally very close to the present work (two of the authors of
the current paper were involved in these experiments): the papers by
{\it Guyer et al.}~[1999] (referred to as GTJ below) and {\it Smith
and TenCate}~[2000] (referred to as S\&T below).  We now address the
question why such differing conclusions were arrived at earlier: was
it the experimental data themselves or were they analyzed and
interpreted incorrectly?  In order to provide the answer we reanalyze
a subset of the older data sets investigated in GTJ and S\&T.

The experiment underlying the two papers was carried out over a long
span of time.  The data set analyzed in GTJ is in fact a small subset
of the data investigated in S\&T, as stated in the second paper
explictly.  The sample under consideration was a Berea sandstone rod,
35~cm long and 2.4~cm in diameter (the numbers quoted in GTJ are
slightly different: 30~cm length and 6~cm diameter, we verified that
S\&T were correct), therefore very similar to the sample used in this
paper.  In order to reduce effects from moisture contained in the
sandstone the sample was kept under vacuum for an extended period.
This increased the quality factor of the rod to $Q\sim 300$ making the
analysis of the experiment easier, since the resonance curves are less
broad than for lower $Q$.  (In GTJ the quality factor is incorrectly
quoted to be $Q=170$, the discrepancy arising due to measuring $Q$
from the width of the resonance curve at half-maximum of the amplitude
rather than at $1/\sqrt{2}$ of the maximum.)  The quality factor in
the old experiment was therefore roughly five times higher than in the
current one.  The resonance frequency in the old experiment was $f\sim
2880$~Hz, which is close the resonance frequency of the sample we
investigated, $f\sim 2755$~Hz.  In the old experiments, different
measurements were made at different temperatures, ranging from
35$^\circ$C to 65$^\circ$C, but for each separate measurement the
temperature was controlled to approximately 0.1$^\circ$C. The
experiments were carried out in three different strain ranges: at very
low strain, at medium strain, and at high strain.  We will be more
specific about the strain ranges below.  The main result found by GTJ
was a linear fall-off of the resonance frequency peak with increasing
strain while S\&T concluded from the same experimental data that the
resonance frequency peak fell off first linearly and then
quadratically with increasing strain.

\begin{figure}[t]
\begin{center}  
\leavevmode\includegraphics[width=7cm,height=5.5cm]
{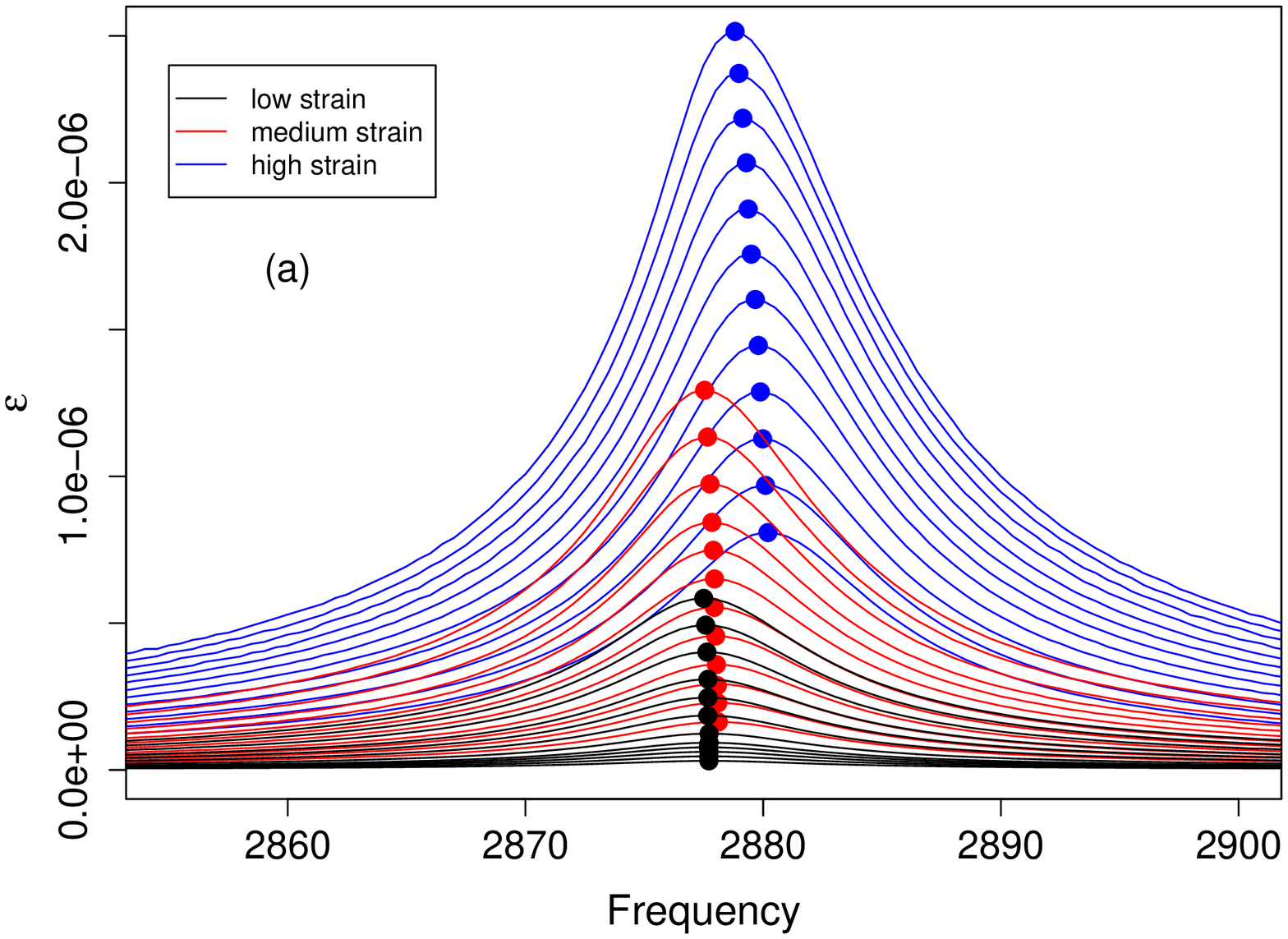} 

\vspace{0.5cm}

\leavevmode\includegraphics[width=7cm,height=5.5cm]
{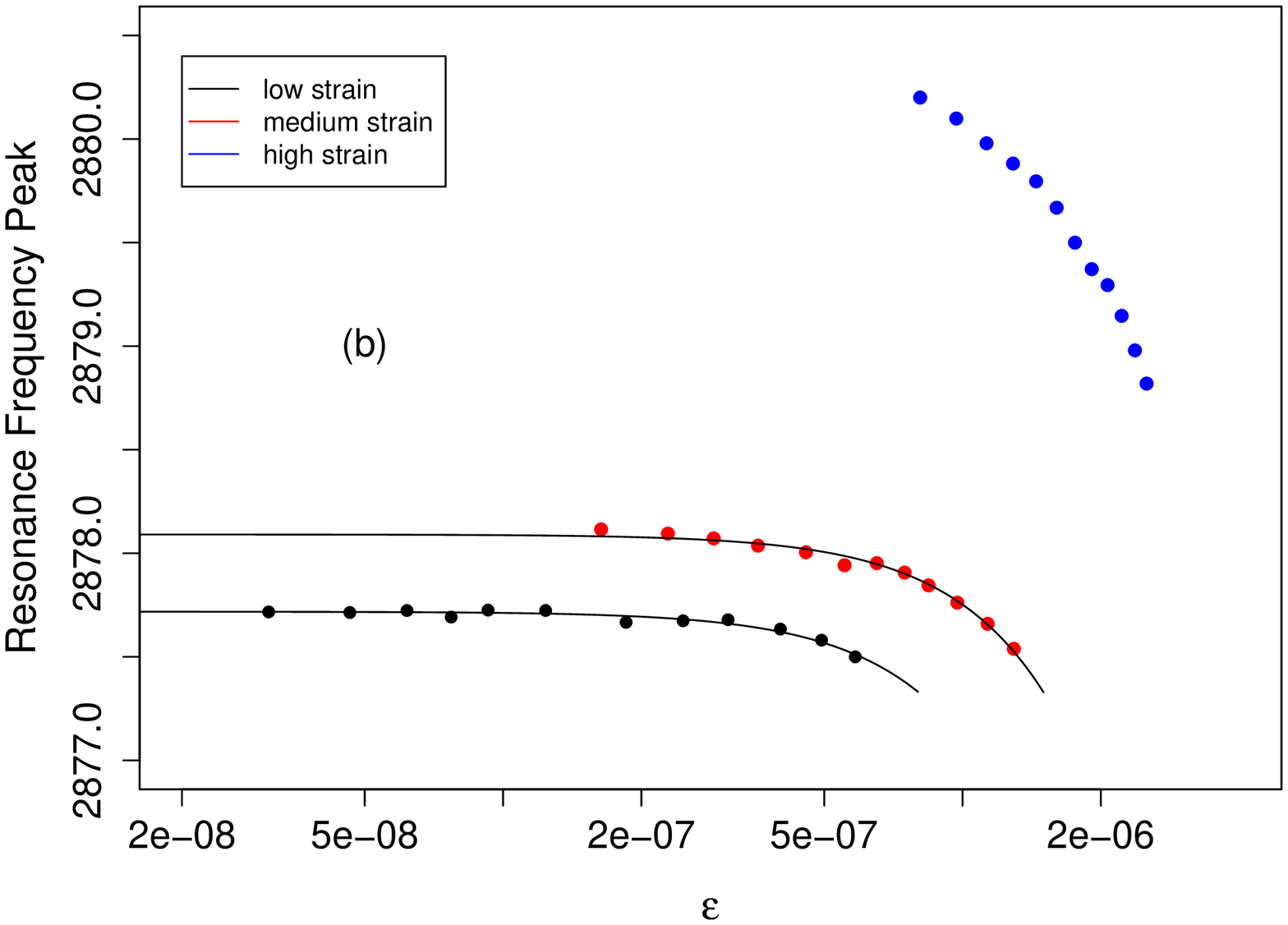}
\end{center}
\caption{Comparison with previous experiments on Berea: (a) resonance
frequency curves for three sets of experiments at three different
strain ranges.  (b) The corresponding resonance frequency peaks.  Note
the logarithmic scale on the $x$-axis. The solid lines represent
predictions of the theoretical model, see Eqn.~(\ref{sig}).}
\label{fig:old} 
\end{figure}

Before we turn to discuss the analysis strategies followed in GTJ and
S\&T we first investigate a subset of the old data set in exactly the
same way as in the new experiments.  The results are shown in
Figure~\ref{fig:old}.  We randomly chose one data set taken at a
constant temperature of 35$^\circ$C. Figure~\ref{fig:old}(a) shows
three sets of resonance curves at different strain ranges.  The peaks
of the resonance curves are determined with our MCMC analysis method
as described in Section~\ref{section3} and marked by the filled
circles.  In Figure~\ref{fig:old}(b) the peaks of the resonance curves
are plotted versus the strain.  From this figure the strain ranges can
be read off: the low strain regime ranges from 3.1$\cdot$10$^{-8}$ to
5.8$\cdot 10^{-7}$, the medium strain regime from 1.64$\cdot 10^{-7}$
to 1.3$\cdot 10^{-6}$, and the high strain regime from 8.1$\cdot
10^{-7}$ to 2.5$\cdot 10^{-6}$.  The solid lines in the low and medium
strain regime represent the predictions from our model.  In these two
regimes the predictions from the Duffing model are excellent, and no
unexpected behavior, such as a linear fall-off is observed.  Note,
that the model in this case works even at higher strains than the
threshold found in the new experiment, although of course $\epsilon_M$
in the old experimental samples could have been different.  The
measurements in the high strain regime are contaminated by
nonequilibrium effects and therefore our simple model is not
applicable.  To reiterate: the old data set reanalyzed by us is in
complete agreement with the results from our new experiments.  A
threshold where the Berea sample behaves as a linear material exists,
for low strains the sample behaves like a classical nonlinear
material, and at very high strain, due to nonequilibrium effects, the
interpretation of the data becomes very involved and does not allow
for deciding between classical or nonclassical behavior.

After verifying that the old experimental data in no way contradict the
results from our new experiments we now turn to the analysis
strategies used in GTJ and S\&T and the interpretations of their
findings.

In contrast to our analysis, in which we determine the peak of every
single resonant curve, GTJ analyze the data at constant strain.  While
this method should work in general, it has several shortcomings.
First, the resonance curves analyzed in GTJ were only sparsely sampled
with data points.  In order to carry out the constant strain analysis,
the resonance curves had to be interpolated to obtain the values at
one constant strain.  This fitting procedure might lead to a bias in
the results with respect to the functional form of the fit applied.
Second, the number of data points available for the application of the
constant strain analysis decreases rapidly with strain amplitude.
Third, the constant strain analysis leads to correlated error bars. 
(Our MCMC-based method is free from these problems.)  

The results for the dependence of the resonance frequency versus
strain are shown in Figure~3(a) in GTJ. The strain range shown on the
$x$-axis in this plot is $10^{-8}$ to 5$\cdot 10^{-7}$ as explained in
the text.  The three different curves GTJ show are from different
measurements and in all cases the dynamic range is very small.
Consider now the lowest -- and longest -- of these curves, the strain
range here is only 10$^{-7}$ to $3\cdot 10^{-7}$. A linear fit for
this data set might naively appear to be justified, even though the
data points at the higher strains are already falling off a linear
fit.

\begin{figure}
\begin{center}  
\leavevmode\includegraphics[width=7cm,height=5.5cm]
{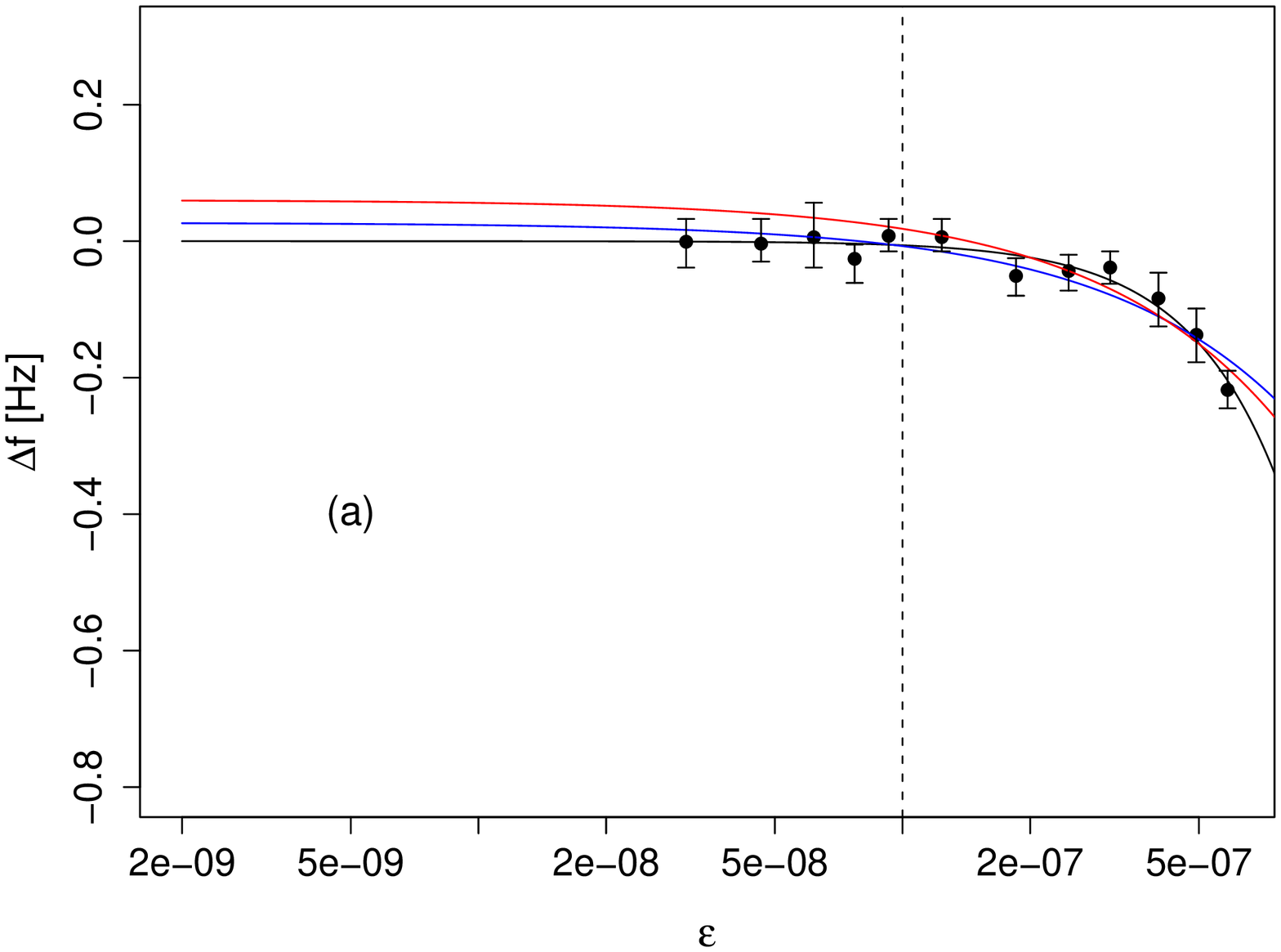} 

\vspace{0.5cm}

\leavevmode\includegraphics[width=7cm,height=5.5cm]
{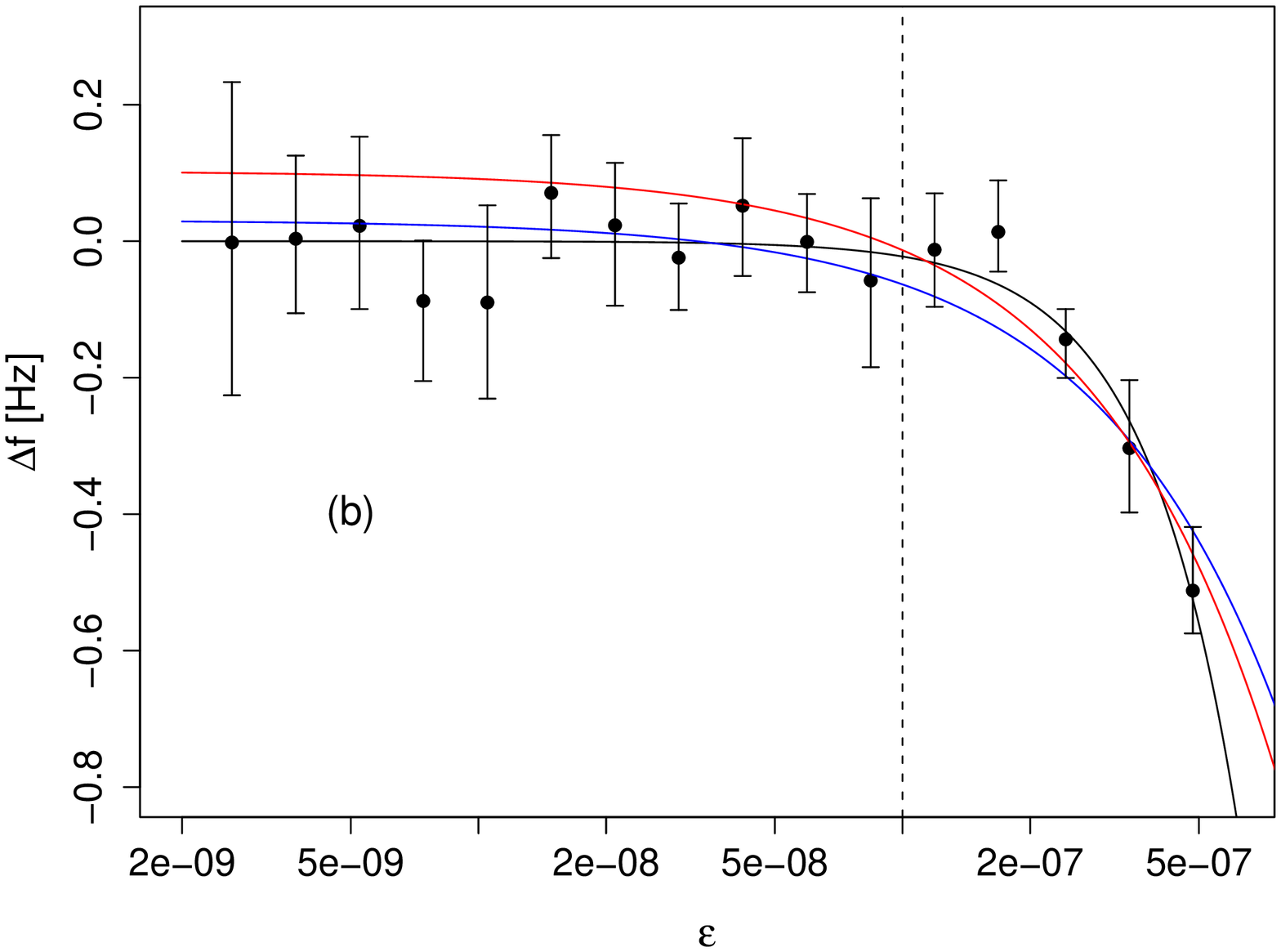}
\end{center}
\caption{Comparison of different fits for (a) the old Berea data set
and (b) the new Berea data set.  Note the logarithmic scale on the
$x$-axis.  The black line shows the quadratic fit obtained from the
Duffing model, the red line shows the best linear fit including data
points only to the right of the dashed line.  Inclusion of all data
points for the linear fit makes the fit much worse.}
\label{fig:oldfit} 
\end{figure}

To emphasize the importance of having sufficient dynamic range, we
return to Figure~\ref{fig:old}(b) and consider only the lowest strain
measurement data set, shown in detail in Figure~\ref{fig:oldfit}(a).
The dashed line marks the strain corresponding to the lowest strain in
GTJ in their longest strain range measurement.  We show in red the
best linear fit to all the data points on the right of this line.  The
highest strain in GTJ was $3\cdot 10^{-7}$ so would only include 4 of
the data points in Figure~\ref{fig:oldfit}(a).  If we only concentrate
on the strain regime to the right of the dashed line, both fits,
linear and quadratic are acceptable.  But if we consider all the
available data points down to the lowest strain, the linear fit fails
by being too high.  Therefore, in order to make a definite statement
about the best fit to the data it is clearly important to have a
sufficient range in strain.

It is not possible to obtain uncontaminated measurements at higher
strains as discussed in detail earlier, hence extension of dynamic
range requires measurements at low strains, as carried out in the
present work. We demonstrate the usefulness of this in
Figure~\ref{fig:oldfit}(b) where we show once again the new Berea
measurements with the quadratic fit shown in black, and the best
linear fit -- again only for the data points on the right side of the
dashed line -- in red.  Inclusion of more points for the linear fit
again makes the agreement much worse. It is apparent that without
sufficient dynamic range it is easy to be misled in fitting a linear
curve to the data.  Including all the data points down to a strain of
10$^{-9}$ demonstrates the correctness of the quadratic fit.  To
summarize: the experimental data in GTJ is apparently correct, but the
dynamic range of the data points analyzed is {\it not} sufficient to
draw any conclusion regarding the nonlinear behavior of the material.

Finally we are unable to understand the remark in GTJ that the
traditional theory of nonlinear elasticity predicts a value of $\Delta
f/f_0 \sim 10^{-10}$ at a strain of roughly $3\cdot 10^{-7}$. All
that traditional theory predicts is a quadratic frequency shift which
we do observe; the magnitude is set by a certain dynamic nonlinearity
coefficient which, in effect, is measured in the experiment. No
contradiction with classical nonlinear theory is observed in our
experiment or indeed in the data of GTJ.

Next, we turn to the results found in S\&T. One of the main objectives
in that work was to investigate the dependence of the frequency shift
(hence, the shift in the Young's modulus) as a function of temperature
changes. The idea was that static hysteresis mechanisms (if present at
very low strains) could be due to thermal activation instead of
mechanical stick-slip processes as in quasi-static experiments at much
higher strain.  (However, the authors did not directly investigate if
the system showed cuspy hysteretic behavior in the first place.)

Experiments at temperatures ranging from 35$^\circ$ to 65$^\circ$ were
carried out.  In addition different strain regimes were investigated
at different times, as shown in the previous Figure \ref{fig:old}(a).
The condition of the rock might have changed in between these
different times, which could have led to a contamination of the
results. Data-fitting was carried out by fitting to a simple pole
response characteristic, however, the possible systematic errors in
this procedure were not discussed. In addition, for each temperature,
the three different sets of measurements at low, medium, and high
strain were shifted in order to obtain a single measurement over a
wide strain range. This approach is likely to lead to a bias in the
result since the rock might have been in different metastable
conditioned states for each data set.

In the final step, the relative shift in the Young's modulus was
determined and fitted by a single function for {\it all} resonance
frequency shift curves, independent of the temperature at which they
were taken or the resonance frequency $\omega_0$ (recall the different
strain ranges of the data shown in Figure~\ref{fig:old}).  The result
of this analysis is shown in Figure~6 of S\&T. It is immediately clear
that a single fit to all the curves is rather dangerous and the single
fit (solid line in Figure~6) does not work particularly well.  The
functional form of the fit -- first linear and then quadratic -- is
therefore also not very meaningful, since no error bars are shown, any
other functional form, such as pure quadratic, would have probably
worked as well.

The authors' contention that the temperature-insensitivity of the
coefficients determining the frequency shift is directly related to
the underlying loss mechanism -- and hence rules out thermal
activation mechanisms -- is incorrect. The relationship between the
frequency shift and the loss mechanism is yet to be elucidated: as
shown in the present work for example, nonlinear frequency shifts and
linear losses can easily coexist and it is well-known that the loss
factor is temperature-dependent.

In summary, the measurements used in GTJ and S\&T are in fact in very
good agreement with our current measurements and understanding of the
nonlinearities in rocks below a certain strain threshold -- it is the
interpretation of the data in these two papers that must be
corrected. In GTJ the strain range over which the analysis was carried
out was insufficient to reach any conclusive result about the fall-off
of the resonance frequency peak with strain. In S\&T the fitting
procedure applied to the data sets seems to have led to erroneous
conclusions about the behavior of $\Delta f$ versus strain.
 
\section{Summary and Outlook} 
\label{conclusions}
In this paper we have described a set of resonant bar experiments
carried out for Berea, Fontainebleau, and Acrylic (as a linear control
material) in order to investigate the dynamic compliance and loss
mechanisms at low strains, between $5\cdot 10^{-8}$ and $2\cdot
10^{-6}$.  To ensure isolation from environmental influences, such as
temperature and humidity, an isolation chamber was employed to obtain
controlled and repeatable results.

The main conclusion of our work is the demarcation of two strain
regimes: in the first regime the material displays reversible
softening of the resonance frequency, while in the second regime,
which occurs after a material and environment-dependent threshold
$\epsilon_M$, nonequilibrium and conditioning effects become
important. Some of these results were previously reported in a short
communication [{\it TenCate et al.}~2004]. Here we report the results
of a detailed study for the first strain regime -- below $\epsilon_M$
-- for both Berea and Fontainebleau samples measuring quantities such
as the quality factor, stress-strain loops, and amplitudes of higher
harmonics.  By repeating measurements on the same samples we have
demonstrated the robustness of the results. At strains characteristic
of reversible nonlinear behavior, the quality factor is essentially
constant, but it is possible that it reduces at higher strain
values. It is not unreasonable to speculate that -- unlike the
resonance frequency shift -- the amplitude dependence of the quality
factor is connected to the onset of nonequilibrium behavior, but this
aspect requires further investigation.

The data analysis was carried using a statistical method based on a
Gaussian process model.  This parameter-free method avoids any biasing
of the analysis due to fitting of the resonance curves with specific
functional forms. It also determines reliable error bars for the
resonance frequency shift $\Delta f$ as a function of the applied
drive strength. The vast majority of previous papers analyzing similar
experiments do not provide a detailed error analysis.

A theoretical framework for the experimental results is provided by a
simple damped Duffing model for which closed-form results can be
obtained. The Duffing model predictions are in excellent agreement
with the entire set of experimental measurements over the strain
regime $\epsilon < \epsilon_M$.

Our results are in disagreement with some of the previous work carried
out with the resonant bar technique as has been pointed out at the
relevant places in the main body of the paper. In two cases -- {\em
Smith and TenCate}, [2000] and {\em Guyer et al.},~[1999] -- we have
reanalyzed a subset of the older experimental data and have
demonstrated that the disagreement is {\em not} due to fundamental
differences in the data but due to mistakes in the theoretical
interpretation and analysis in these papers. Thus, one goal of this
paper is simply to clarify the present state of knowledge in the
low-strain regime.
 
While in this paper, we have focused on the reversible nonlinear
regime ($\epsilon < \epsilon_M$), future work will target the
understanding of the nonequilibrium behavior of geomaterials. The
investigation of this second regime is at the same time fascinating
and very challenging. It is difficult, but essential, to disentangle
conditioning/nonequilibrium and nonlinear effects. New experimental
strategies have to be developed for this endeavor. At the same time a
theoretical framework which encompasses and explains all known
physical effects needs to be developed.

\begin{acknowledgments}
This work was funded in part through the DOE Office of Basic Energy
Science and the Institute of Geophysics and Planetary Physics of Los
Alamos National Laboratory (IGPP).
\end{acknowledgments}

%------------------------------------ BIBLIOGRAPHY--------------------

\end{article}
\end{document}